\documentclass{sig-alternate-05-2015}

\usepackage{times} 

\usepackage{amsmath}
\usepackage{amsfonts}
\usepackage{amssymb}
\usepackage{graphicx}
\usepackage{color}
\usepackage{url}
\usepackage[boxed,vlined]{algorithm2e}
\usepackage{paralist} 
\usepackage{listings} 

\usepackage{subfigure}
\usepackage{booktabs}
\usepackage{multirow}
\usepackage[table]{xcolor}
\usepackage{enumitem}
\usepackage{wasysym}
\usepackage{pifont}
\usepackage{bbm}

\begin{document}

\setcopyright{acmcopyright}


\newtheorem{observation}{Observation}
\newtheorem{conjecture}{Conjecture}
\newtheorem{problem}{Problem}
\newtheorem{algo}{Algorithm}
\newtheorem{lemma}{Lemma}
\newtheorem{question}{Question}
\newtheorem{example}{Example}
\newtheorem{answer}{Answer}

\newcommand{\hide}[1]{}
\newcommand{\notice}[1]{{\textsf{\textcolor{green}{{\em [#1]}}}}}
\newcommand{\reminder}[1]{{\textsf{\textcolor{red}{[#1]}}}}
\newcommand{\vectornorm}[1]{\left|\left|#1\right|\right|}

\newcommand{\mkclean}{
    \renewcommand{\reminder}{\hide}
}

\newcommand{\bit}{\begin{compactitem}}
\newcommand{\eit}{\end{compactitem}}
\newcommand{\ben}{\begin{compactenum}}
\newcommand{\een}{\end{compactenum}}

\newcommand*{\QED}{\hfill\ensuremath{\blacksquare}}%
\newcommand{\method}{\textsc{OurMETHOD}\xspace}
\newcommand{\model}{\textsc{NetTide}\xspace}
\newcommand{\modelInTitle}{{\textbf{\large \textsc{NetTide}}}\xspace}
\newcommand{\goal}{\textsc{Goal}\xspace}
\newcommand{\NA}{---}

\newcommand{\douli}{\textsc{Douli}}
\newcommand{\mapreduce}{\textsc{MapReduce}\xspace}
\newcommand{\hadoop}{\textsc{Hadoop}\xspace}
\newcommand{\nodeiterator}{\textsc{NodeIterator}\xspace}
\newcommand{\edgeiterator}{\textsc{EdgeIterator}\xspace}
\newcommand{\doulinodeiterator}{\textsc{DouliNodeIterator}\xspace}
\newcommand{\eigen}{\textsc{EigenTriangle}\xspace}

\newcommand{\nt} {n(t)}
\newcommand{\et}{e(t)}
\newcommand {\Indicator}{\mathbbm{1}}

\newcommand{\qAccuracy}{Q1 - Accuracy}

\newcommand{\head}[1]{\textnormal{\textbf{#1}}}
\definecolor{TableBorder}{rgb}{0.0078,0.6824,0.7882}
\definecolor{TableEven}{rgb}{0.8000,0.9216,0.9490}
\definecolor{TableOdd}{rgb}{1,1,1}

\definecolor{myred}{RGB}{228, 26, 28}
\definecolor{myblue}{RGB}{55, 126, 184}

\newcommand{\mytag}[1]{ {\bf #1}}

\title{
Structural patterns of information cascades and their implications for dynamics and semantics
}

\numberofauthors{1}
\author{
\alignauthor
Chengxi Zang$^1$, Peng Cui$^1$, Chaoming Song$^2$, Christos Faloutsos$^3$ and Wenwu Zhu$^1$\\
       \affaddr{$^1$
       Department of Computer Science, Tsinghua University, Beijing, China}\\
       \affaddr{$^2$ Department of Physics, University of Miami}\\
       \affaddr{$^3$ Computer Science Department, Carnegie Mellon University}\\
       \email{\normalsize zangcx13@mails.tsinghua.edu.cn, chaomingsong@gmail.com, christos@cs.cmu.edu, \{cuip,wwzhu\}@tsinghua.edu.cn}}

\maketitle
\begin{abstract}

Information cascades are ubiquitous in both physical society and online social media, taking on large variations in structures, dynamics and semantics. Although the dynamics and semantics of information cascades have been studied, the structural patterns and their correlations with dynamics and semantics are largely unknown. Here we explore a large-scale dataset including $432$ million information cascades with explicit records of spreading traces, spreading behaviors, information content as well as user profiles. We find that the structural complexity of information cascades is far beyond the previous conjectures. We first propose a ten-dimensional metric to quantify the structural characteristics of information cascades, reflecting cascade size, silhouette, direction and activity aspects. 
We find that bimodal law governs majority of the metrics, information flows in cascades have four directions, and the self-loop number and average activity of cascades follows power law. 
We then analyze the high-order structural patterns of information cascades.
Finally, we evaluate to what extent the structural features of information cascades can explain its dynamic patterns and semantics, and finally uncover some notable implications of structural patterns in information cascades. Our discoveries also provide a foundation for the microscopic mechanisms for information spreading, potentially leading to implications for cascade prediction and outlier detection.

\end{abstract}


\section{Introduction}
\label{sec:intro}
Information spreading is a ubiquitous phenomenon in self-organized social systems, enabling the local individuals to have global senses, and thus playing important roles in news propagation \cite{ye2010measuring}, innovative technology dissemination \cite{rogers2010diffusion}, as well as epidemic diffusion \cite{pastor2015epidemic}. Due to the importance and complexity of this phenomenon, the generated information cascades have attracted considerable attention in recent years, ranging from the cascades of chain-letters \cite{liben2008tracing} in physical society to the cascades of resharing in on-line social media platforms such as Facebook \cite{cheng2014can}, Twitter \cite{romero2011differences}, and Weibo \cite{yu2015micro}. Although the dynamics \cite{yang2011patterns,matsubara2012rise,Zang2016} and semantics \cite{romero2011differences,yang2014large} of information cascades are explored, \hide{the structural patterns of information cascades are largely unknown. I}it remains an interesting problem to see how to quantify the structural patterns of information cascades, and whether the structural patterns have notable correlations with dynamics and semantics. The answer to this question is of paramount significance for both uncovering the intrinsic mechanism of information spreading in scientific research and imposing good forecasting and controlling over information cascades in real applications.

The major reason why this problem is rarely addressed is the lack of datasets which explicitly record the full traces of information flow.
In this paper, we collect $432$ million information cascades in Tencent Weibo \footnote{http://t.qq.com/}, which is one of the largest microblog systems in China. This dataset covers \emph{ the full scale} information cascades generated during one week. For each microblog, we have the explicit records of its spreading traces, the timestamps of spreading behaviors, the content, as well as the profiles of involved users. As far as we know, this is among the first datasets that can support the target study.

Through extensive observational study over the dataset, we find that the structural complexity of information cascades is far beyond our expectation. In order to quantify the complex structural characteristics, we propose a ten-dimensional structural metric to reflect the size, silhouette, direction and activity aspects of information cascades. 
We find: (a) the bimodal distribution governs the mass, length, breadth, wiener index and the number of reciprocal edges;
(b) information flows in empirical cascades have four directions, namely branching-out, converging-in, reciprocal, and self-loop;
(c) repeated retweets and self-promotions are prevalent, and the average activity and the number of self-loops follow power law.
We further study the 
high-order structural patterns of cascades by answering: (a) Are cascade wide and shallow? Or narrow and deep? (b) To what extent do cascades follow the star-like pattern or chain-like pattern? (c) To what extent do the four directions of information flow coexist in a cascade? (d) What are the structure patterns of so-called popular cascades?

Based on the structural analysis of information cascades, we analyze their correlations with the dynamics and semantics. After clustering information cascades into dynamic clusters and topic clusters, we evaluate to what extent can structural features of a information cascade explain its corresponding dynamic cluster and topic cluster. The results show that the structural features of a information cascade have notable correlations with its dynamic and semantic features. Through more insightful case studies, we also provide answers to the following questions: 
(a) Which dynamic/semantic patterns are bigger in mass, longer in length and wider in breadth? (b) Which dynamic/semantic patterns are viral in structure? (c) Which dynamic/semantic patterns show polarized spreading directions? (d) Which dynamic/semantic patterns show borrowed prosperity?

Our discoveries may provide foundations for the intrinsic mechanisms for information spreading, potentially leading to applications including cascade prediction, influence maximization, and outlier detection.

\section{Related Work}
\label{sec:background}
We presented related work in three dimensions: cascade structures, dynamics and topics. 
The cascade structures have been studied in different scenarios, including blogs \cite{adar2004implicit,leskovec2007patterns,pei2015exploring}, communication network \cite{kossinets2008structure}, Facebook \cite{dow2013anatomy}, Twitter \cite{goel2012structure}, LinkedIn \cite{anderson2015global}.\hide{, or to infer structures \cite{gomez2013structure}.} Deep-tree structures are found and modeled in Ref.\cite{liben2008tracing}.
Goel \cite{goel2015structural} studied the virality of cascades by using Wiener index. However, we find information flows in cascades have four directions, indicating complex structures than tree patterns. Furthermore, ignoring the high-order correlations between structure metrics in previous studies prevent our understandings of cascade structures.

The dynamics of cascades have been empirically studied in marketing \cite{leskovec2007dynamics}, Twitter\cite{myers2014bursty}, Facebook \cite{cheng2016cascades}, QQ \cite{zhang2016multiscale}, and so on.  Another line of works try to model the dynamics from differential equation and micro-process \cite{matsubara2012rise,Zang2016}. The prediction of the dynamics are studied in Refs.\cite{cheng2014can,yu2015micro,martin2016exploring}. Time series clustering methods are applied to discover the dynamics patterns or shapes \cite{yang2011patterns,paparrizos2015k}. Our work is based on the clustering method to find out the dynamics clusters in empirical cascade growth dynamics. However, in order to study the correlations between dynamics and structures, we propose new similarity intuitions and find the existing shift-variant and scale-variant algorithm can not do this job.

Previous study shows different topics of information spreading follow different mechanism \cite{romero2011differences}. Celebrated topic modeling method LDA \cite{blei2012probabilistic} is not applicable to Twitter or Weibo scenarios \cite{yang2014large}. In oder to get high precision in topic modeling, we adopt information filtering methods inspired by \cite{hong2010empirical,yang2014large}.

However, to our best knowledge, few of previous works study the interplays between cascade structures, dynamics and topics.

\section{Cascade structure patterns}

\subsection{Experimental setup}
We first give a graph-theoretic definition of cascade and followed by describing our datasets. 

\mytag{Cascade structure definition.}
The structure of a cascade $C=(V,E)$ is a directed graph in which each node $u \in V$ represents a user and each edge $(u,v) \in E$ represents that user $v$ retweets user $u$'s post. The user $u_o \in V$ who initializes the post is the original poster and all the other users are  retweeters. Thus, the depth ($D$) of node $u$ is defined as the number of edges of the shortest path from $u_o$ to $u$, where $u_o$ at depth zero. There exists an integer weight $w(u,v) \geq 1$ which counts the number of multiple edges from $u$ to $v$, indicating the fact that $v$ retweets $u$ $w(u,v)$ times. 
A loop $(u,u)$ is an directed edge that connects $u$ to itself, indicating that user $u$ retweets himself. 
Reciprocal edges $r_{uv}$ are a pair of edges $(u,v) \in E$ and $(v, u)\in E$ where $u \neq v$, indicating the user $u$ and $v$ retweet each other.
 
 
\mytag{Cascade dynamics definition.} Let $C=(V,E,T)$ represent a cascade in which each edge $(u,v,t)\in E \times T$ represents user $u$ retweets $v$ at time $t$ and $\times$ is the Cartesian product of two sets. We define the time of user $u$ being infected $t_u$ as the minimal value of $\{t|(u,v,t)\in E \times T, v \in V \}$. Then, the temporal dynamics of cascade $C$ is the growth rate $c(t)=|\{u \in V| t_u \in [t,t+1)\}|$ where $1$ represents one time unit.  The lifetime of cascade $C$ is the time difference between the posting time of the original poster and the time of the last retweeter. 


\mytag{Dataset.} We get all the posts published in Tencent Weibo spanning from June 20,2012 to June 26, 2012, which involve more than $101$ million users and   $563$ million posts. For each post, we know whether it is an original post or a retweet, the content of the post (including text, embedded hashtag, and urls for embedded pictures and videos), and the timestamp. For each retweet, the dataset records the information source ID. Thanks to the exact records of information pathway, we build more than 432 million information cascades. We also collect the user profiles of these users. For each user, we know the nickname, the tags and  descriptions of the user himself.

\subsection{Quantify cascade structures}
Based on the definition of the cascade structure, we propose  size, silhouette, direction and activity aspects of cascades to quantify their structures.

\subsubsection{Cascade size}
\begin{figure}[!htb]
\centering
\subfigure[Size illustration]{
\includegraphics[width=0.23\textwidth]{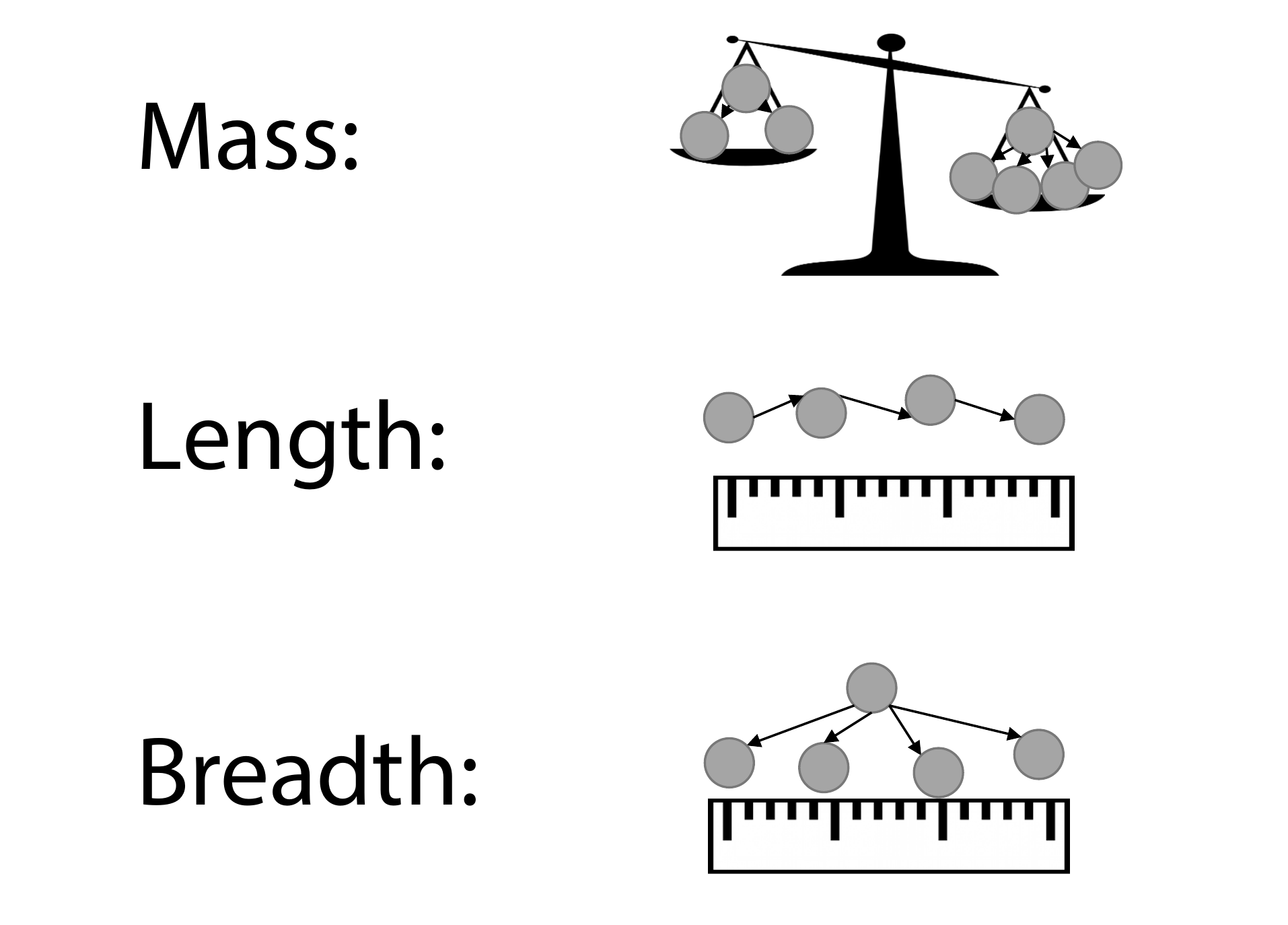}
\label{fig:Intro1}
}
\subfigure[Mass distribution]{
\includegraphics[width=0.22\textwidth]{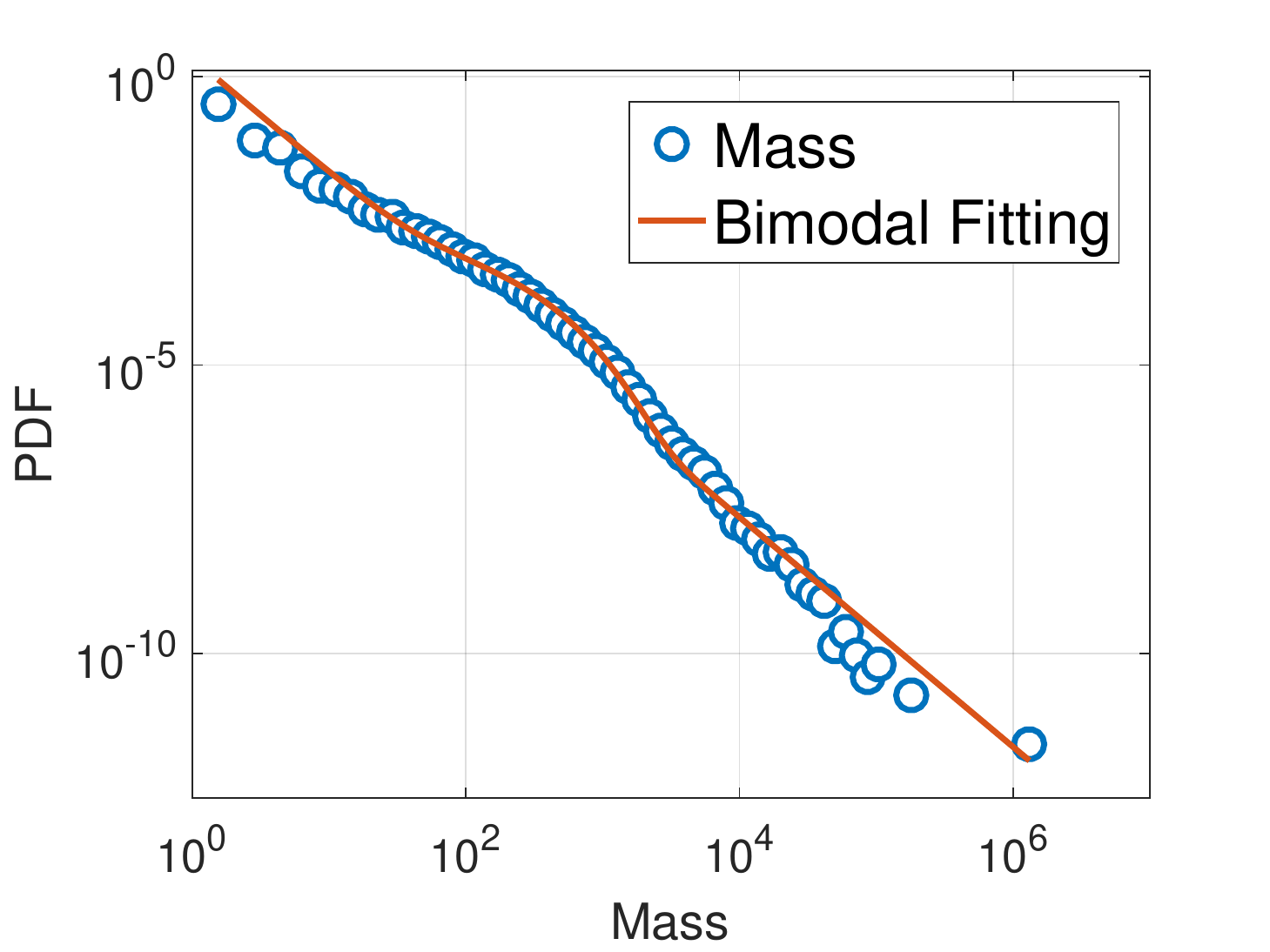}
\label{fig:Intro1}
}
\subfigure[Length distributions]{
\includegraphics[width=0.22\textwidth]{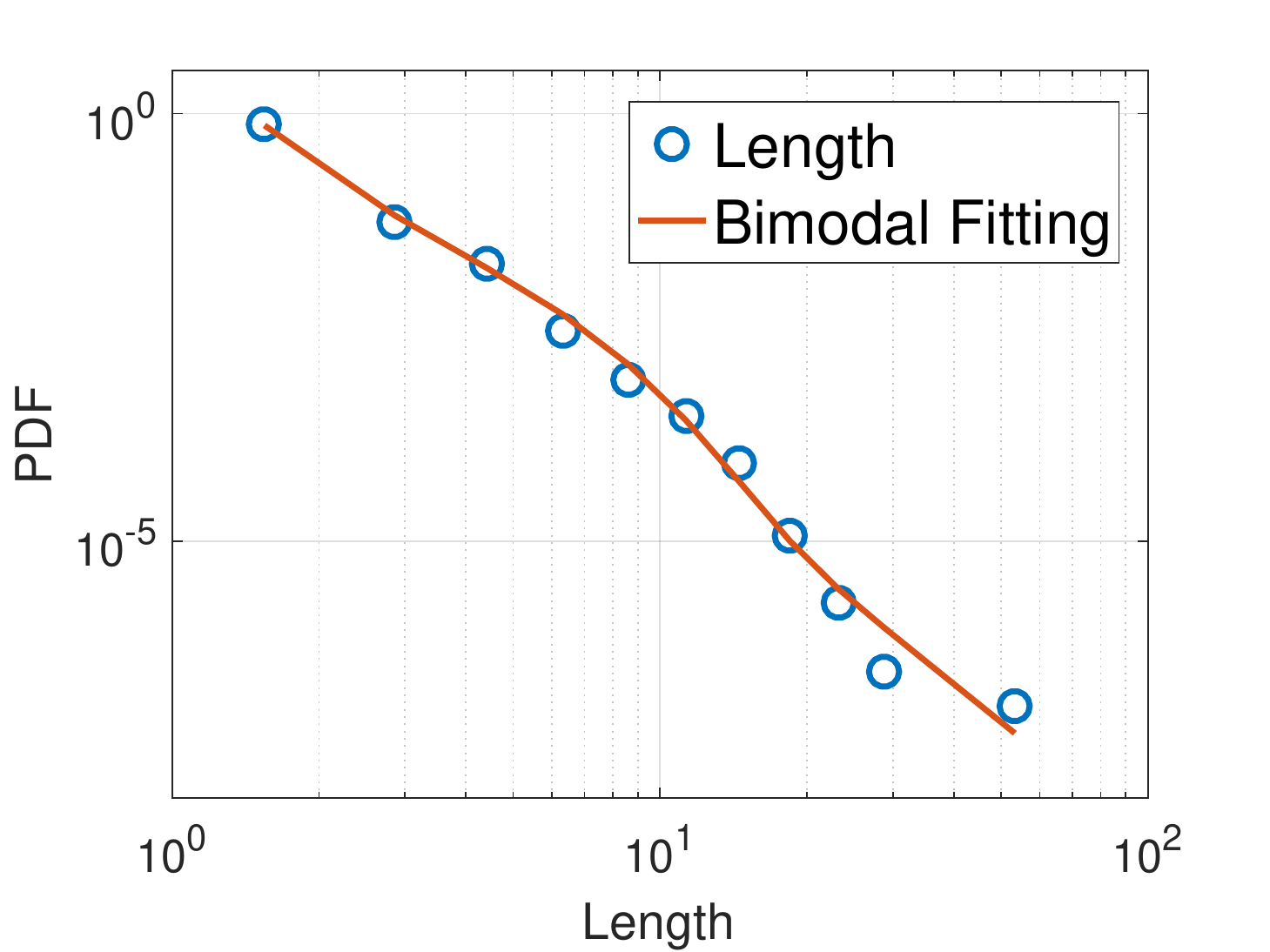}
\label{fig:Intro1}
}
\subfigure[Breadth distribution]{
\includegraphics[width=0.22\textwidth]{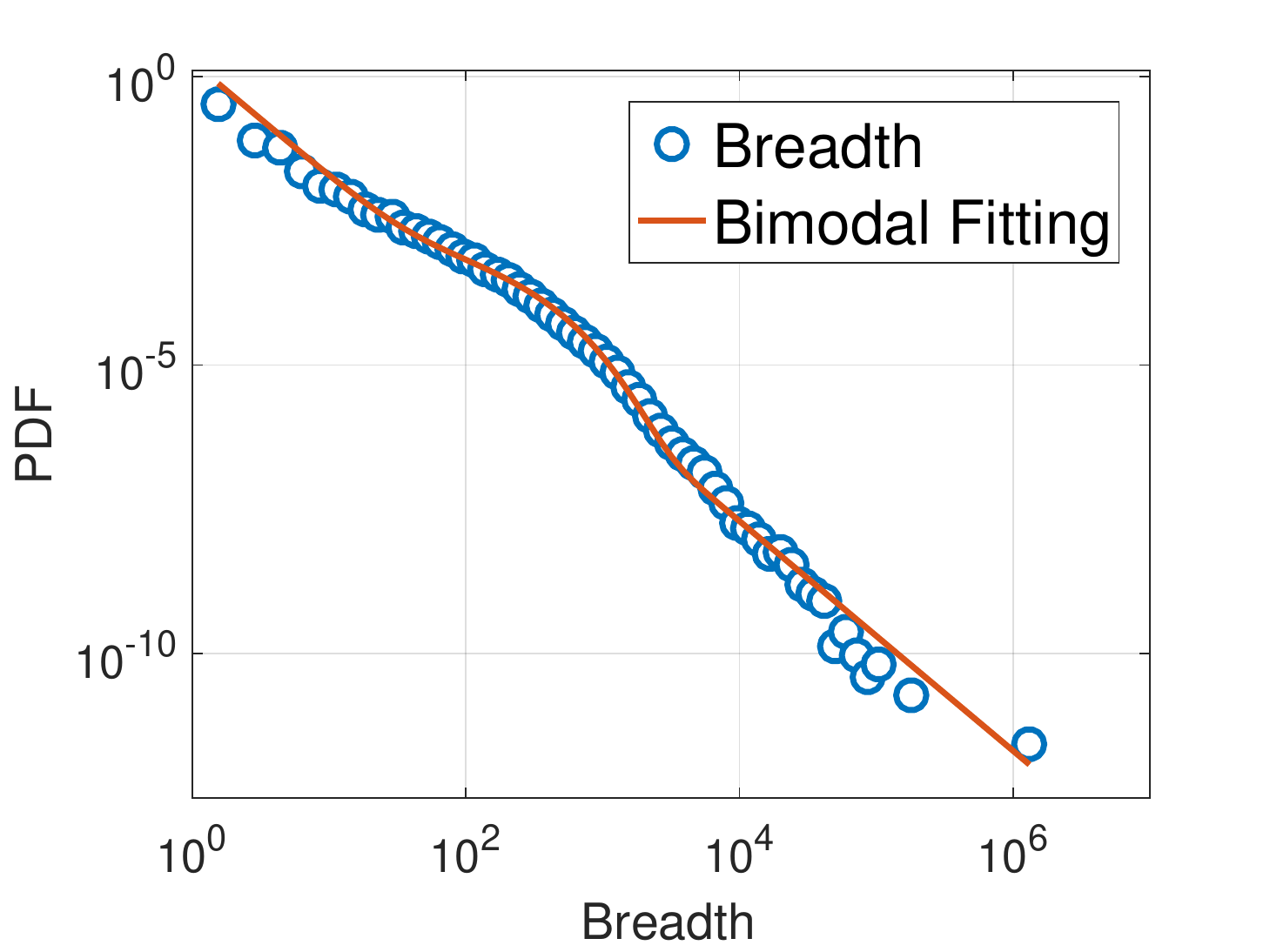}
\label{fig:Intro1}
}
\caption{ \emph{Bimodal size distributions.} (a) Cascade size illustration and (b-d) empirical distributions. We find bimodal distributions for all the size metrics, including (b) mass, (c) length, and (d) breadth. The pdf is binned in the log-log scale. 
\label{fig:size}}
\end{figure} 
 The size concept of a cascades is derived from the need of comparing a bigger to a smaller, a longer to a shorter and a wider to a narrower. Thus, the size of cascade $C$ is measured by following three metrics (as shown in Fig.~\ref{fig:size}a):
\begin{compactitem}
\item \emph{Mass} $N$ of cascade $C$  refers to the amount of unique users in $C$, indicating that a cascade with more users is larger than the one with fewer users. 
\item \emph{Length} $L$ of cascade $C$ is the largest depth,
indicating that a cascade with larger length value is longer than the one with smaller length value.
\item \emph{Breadth} $B$ of cascade $C$  is the largest amount of nodes in $C$ at the same depth, indicating that a cascade with larger breadth value is wider than the one with smaller breadth value. 
\end{compactitem}

\mytag{Fat-tailed and bimodal law for all the size metrics.}
Figures~\ref{fig:size}b-d plot the one-dimensional distribution of the three metrics which quantify the observed cascades size. We observe \emph{fat-tailed} nature for all the three metrics, implying that there exist very large cascades with respect to each size metric. For instance, in our empirical dataset, the biggest cascade which is also the widest cascade has mass value $1,414,815$ and breadth value $1,408,024$. The longest cascade has the length value $57$. However, the average mass, breadth and length values are $5$, $4$ and $1$ respectively. 

Furthermore, we find that all the size metrics, including mass, length, and breadth, exhibit the same \emph{bimodal distribution}:
\begin{equation}
\label{equ:bimodal}
f(x) = c_1 (x + x_0)^{-\alpha} + c_2 e^{-\lambda x^\beta}.
\end{equation}
More specifically, as shown in Figs.~\ref{fig:size}b-d, we find that 
there exist obvious deviation parts from the straight lines in all the log-log plots.
By minimizing the least square error between the Equ.~\ref{equ:bimodal} and the empirical data through the Levenberg-Marquardt algorithm \cite{marquardt1963algorithm}, we get the fitting parameters in Table~\ref{tab:fittingTable} and the fitting curves in Figs.~\ref{fig:size}b-d.
The red fitting curves by Equ.~\ref{equ:bimodal} match the empirical data denoted as circles well, indicating that the probability density function $f(x)$ of all the size metrics follow the mixture of power-law distribution and the stretched exponential distribution. The bimodal distributions are further illustrated in the joint distributions as shown in Fig.~\ref{fig:2dsize}. We will elaborate these findings in the  Sec. 4 High-order structure patterns later.

\begin{table}[!t] 
\label{table:bimodal}
\scriptsize
\caption{The fitting results for the metric distribution by the bimodal distribution $f(x) = c_1 (x + x_0)^{-\alpha} + c_2 e^{-\lambda x^\beta}$.  \label{tab:fittingTable}}
\centering
\begin{tabular}{lcccccc}
\toprule
\textbf{Bimodal}	&\textbf{$c_1$}    &\textbf{$x_0$}  		&\textbf{$\alpha$}  &\textbf{$c_2$} & \textbf{$\lambda$} & \textbf{$\beta$}\\  
\cmidrule{2-7}
Mass  	&2.10 &2.29e-6	&1.99			&1.46e-3 	&0.06   &0.63\\
Length	&5    &1.70e-7  &4.60			&0.11 	    &0.50   &1.05\\
Breadth	&1.80 &2.29e-6	&1.99			&1.46e-3 	&0.06   &0.63\\
Trend   &21	&0.10 &6.43 &0.20 &0.80 & 1.15\\
\#Reciprocal Edge &0.30 &0.01 &2.64 & 2.4e-4 & 0.28 &0.59 \\
\#Self-loop & 3.5e-2 & 0.01 & 2.47 & 0 &- &-\\
Avg-activity &0.79	&0.50 &3.46 &0 &- &-\\
\midrule
\bottomrule
\end{tabular}
\end{table}

\subsubsection{Cascade silhouette}
\begin{figure}[!htb]
\centering
\subfigure[Silhouette illustration]{
\includegraphics[width=0.22\textwidth, trim = 10 0 0 0, clip]{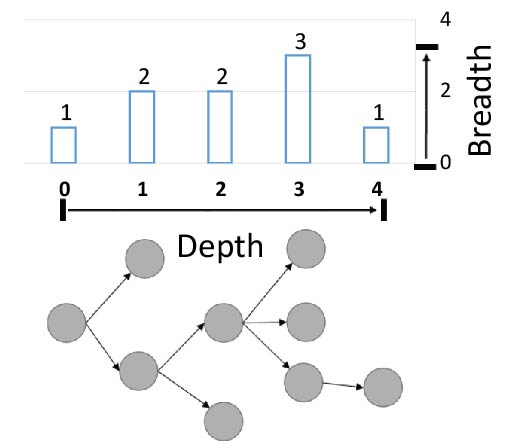}
\label{fig:Intro1}
}
\subfigure[Silhouette distributions]{
\includegraphics[width=0.22\textwidth, trim = 0 0 0 0, clip]{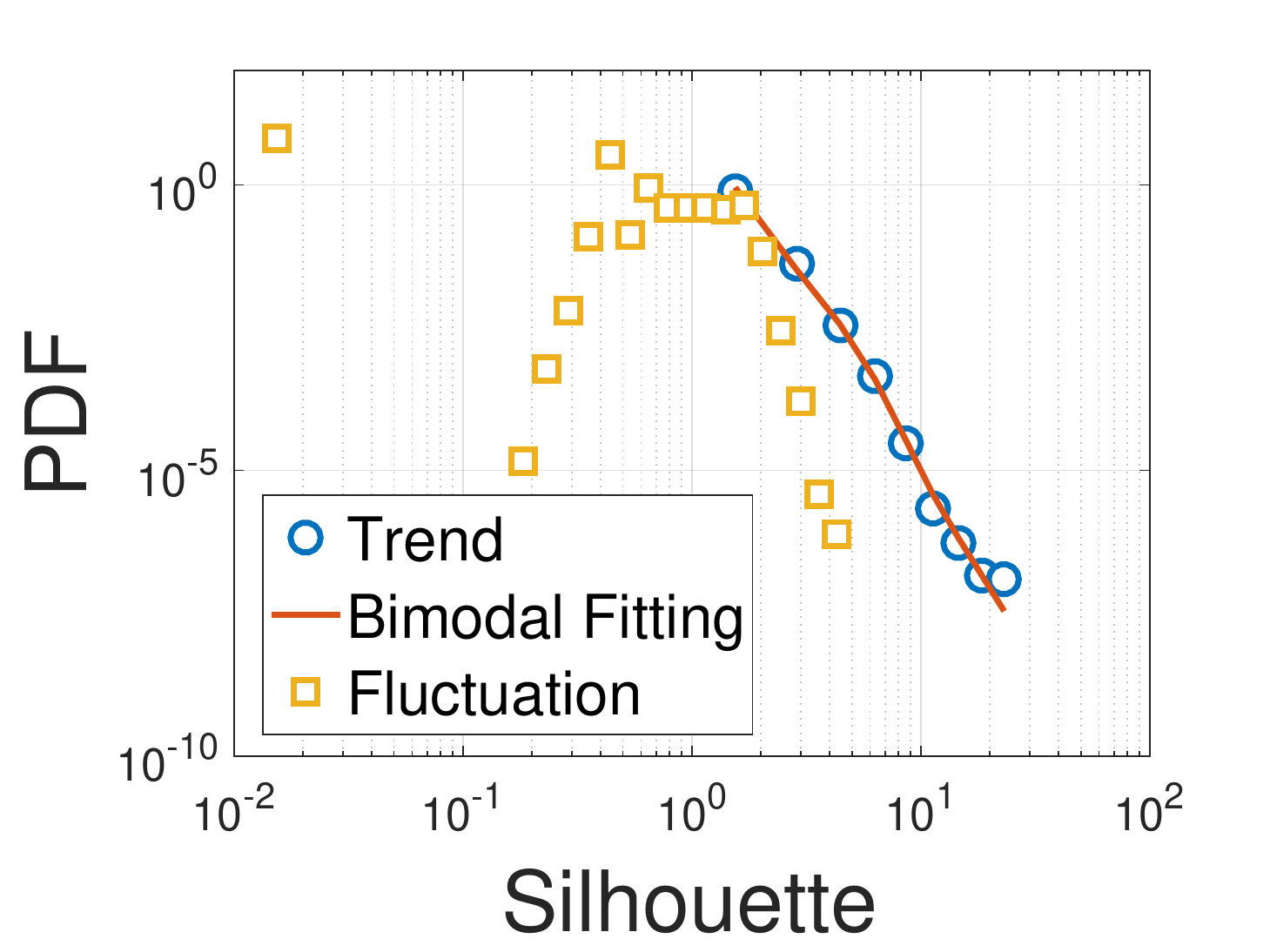}
\label{fig:Intro1}
}
\caption{ \emph{Silhouette of cascades.}(a) Cascade Silhouette illustration and (b) empirical distributions.
\label{fig:shape1d}}
\end{figure}

The silhouette concept characterizes the outline of a cascade.
In order to quantify the silhouette of cascade $C$, we specify each node an integeral coordinate in the two-dimensional Cartesian coordinate system $X \perp Y$ where X-axis and Y-axis represent the depth dimension and breadth dimension of $C$ respectively as shown in  Fig.~\ref{fig:shape1d}a. 
Without loss of generality, the original poster $u_o$ is located at the point $(0, 0)$.  We define $B(D)$ as the breadth value at specific depth $D$, where $D=0,1,...,L$. The $u_o$ is at depth $D=0$, and
thus $B(0) = 1$.
We define the silhouette $S_C$ of the cascade $C$ as the breadth histogram along the depth, namely
$S_C = B(D)$  where $D = 0,1,...,L$ as shown in  Fig.~\ref{fig:shape1d}a. 

To summarize the characteristics of cascade silhouette, we care about whether the silhouette function $S_C=B(D)$ is increasing or decreasing (trend) as $D$ grows, and the relative scale of the extreme points, smoothed or serrated (fluctuation):
\begin{compactitem}
\item \emph{Silhouette trend} describes whether the growth trend of $B(D)$ is increasing or decreasing 
as the depth $D$ increases. Mathematically, the trend is characterized by the first order derivative of $B(D)$. In order to summarize this trend, approximately, we use the Wiener index introduced in Ref. \cite{goel2015structural} which is defined as the average shortest distance between each pair of nodes in $C$ (we treat each directed edge in $C$ as a bidirectional edge because so far we can ignore their connections). 
A cascade $C$ with an increasing value of $B(D)$ as $D$ grows, has a larger Wiener index value because the shortest path between any two nodes at large depth tend to take the nodes with small depth as intermediate points. In contrast, a $C$ with large $B(D)$ value at a small depth $D$ has small trend or Wiener index value. 

\item \emph{Silhouette fluctuation} captures the degree of fluctuations of the outline of $S_C$, characterized by the coefficient of variation of $B(D), D=0,1,...,L$, namely $\frac{std(B(D))}{mean(B(D))}$. For instance, a narrow structure  is characterized by a small fluctuation value, while a fanning-out structure or a wide-and-deep structure have a large fluctuation value.
\end{compactitem}

\mytag{Bimodal law for silhouette metrics.} We find the cascade trend value and fluctuation value also follow bimodal distribution as shown in Fig.~\ref{fig:shape1d}b. Specifically, the trend value follows the bimodal distribution indicated by Equ.~\ref{equ:bimodal}, while fluctuation value follows a more complex bimodal distribution.

\subsubsection{Cascade direction}
\begin{figure}[!htb]
\centering
\subfigure[Venn diagram of converging-in, reciprocal and self-loop]{
\includegraphics[width=0.23\textwidth, trim = 0 0 0 0, clip]{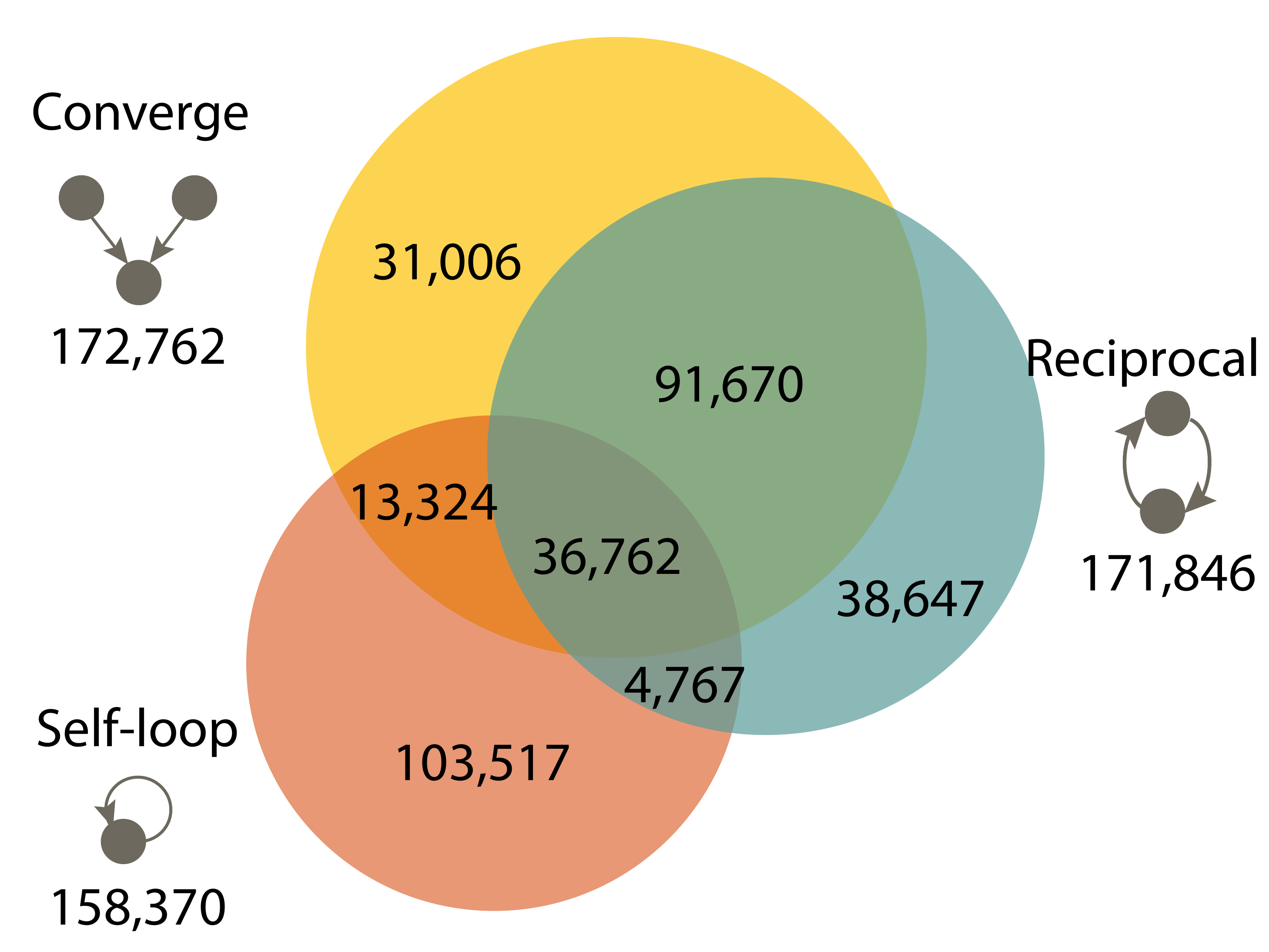}
\label{fig:venn}
}
\subfigure[Branch \& Converge]{
\includegraphics[width=0.21\textwidth, trim = 0 0 0 0, clip]{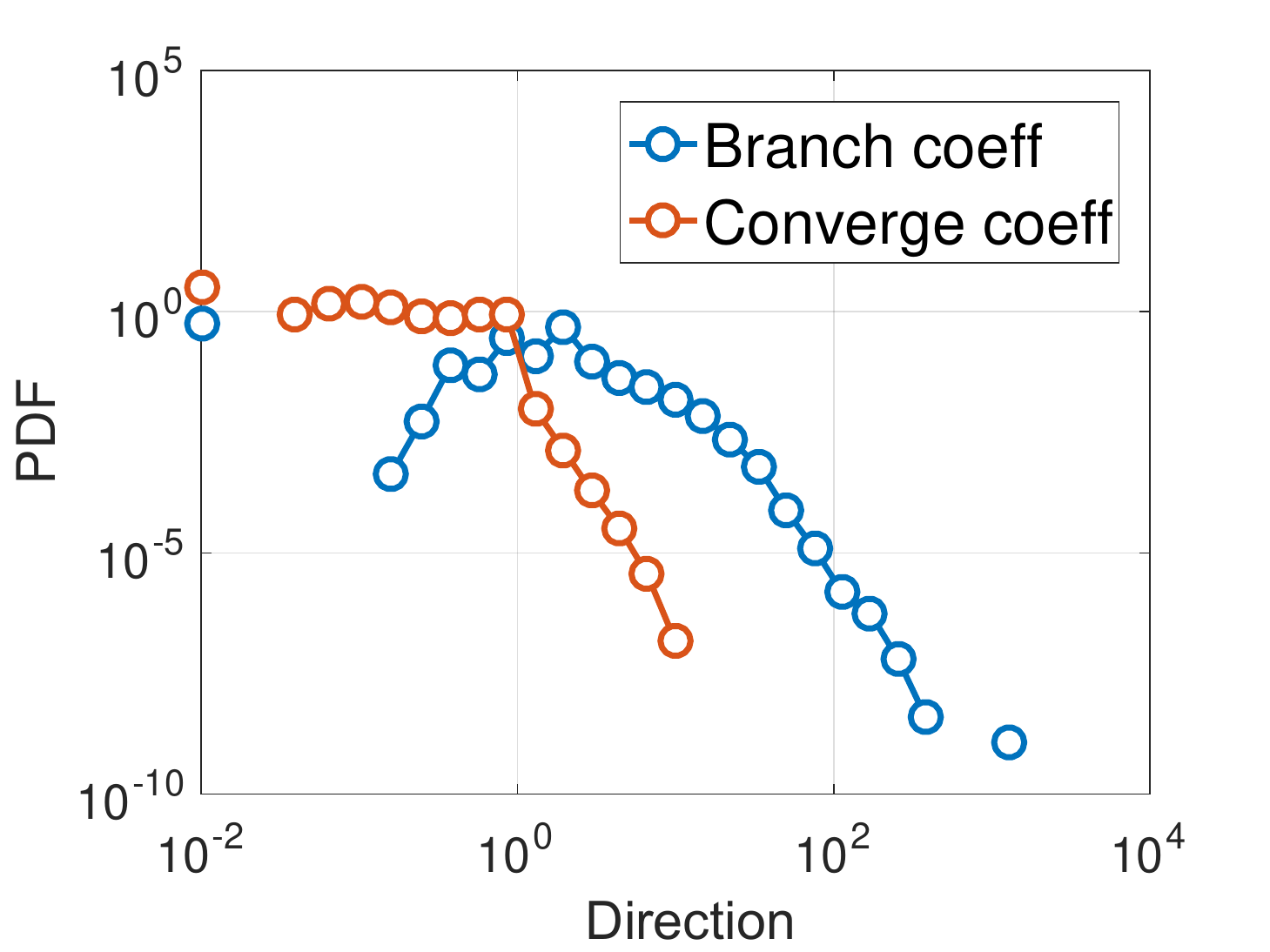}
\label{fig:Intro1}
}
\subfigure[\#Reciprocal \& \#Self-Loop]{
\includegraphics[width=0.23\textwidth, trim = 0 0 0 0, clip]{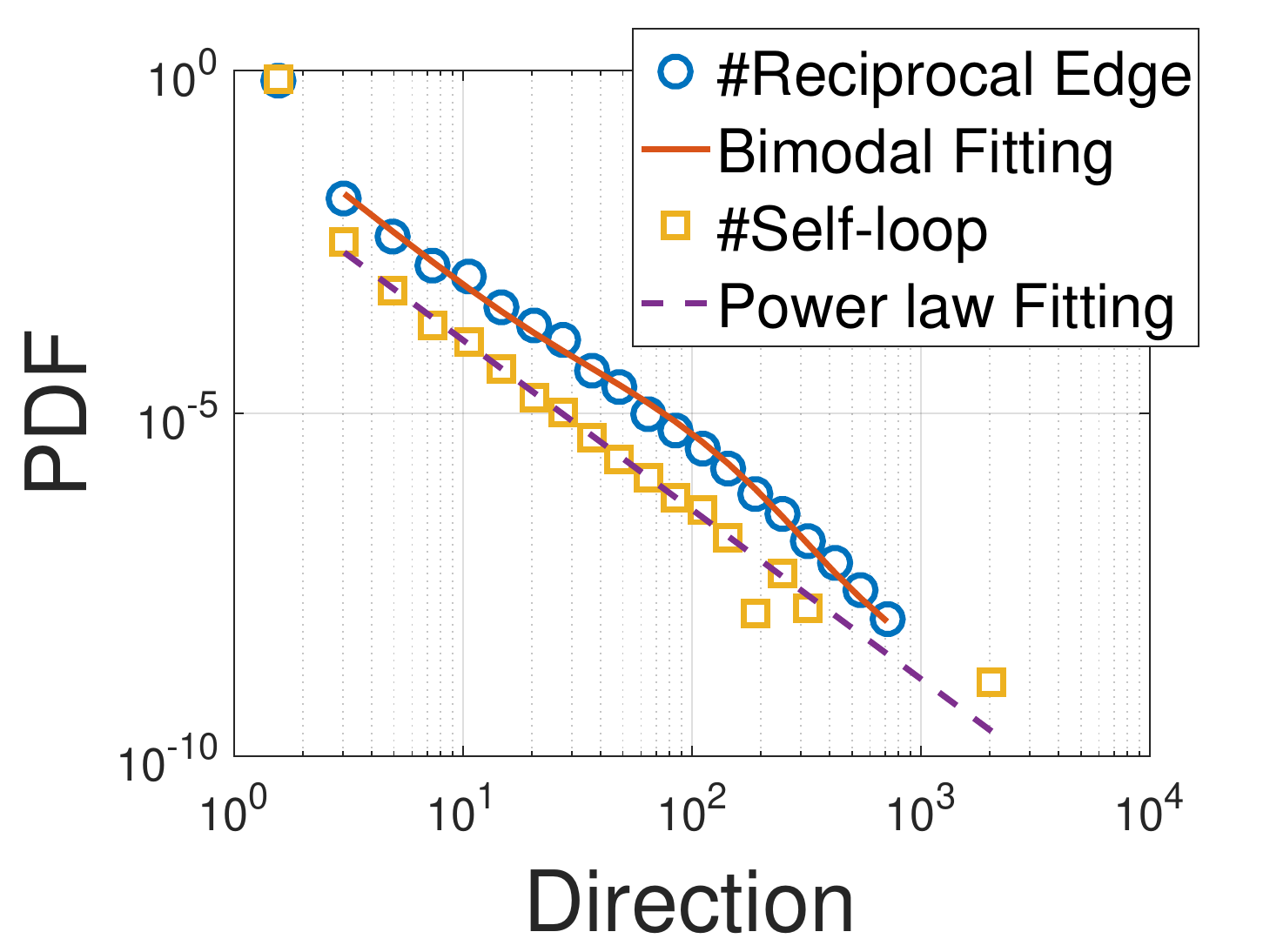}
\label{fig:Intro2}
}
\subfigure[Reciprocal \& Self-Loop]{
\includegraphics[width=0.21\textwidth, trim = 0 0 0 0, clip]{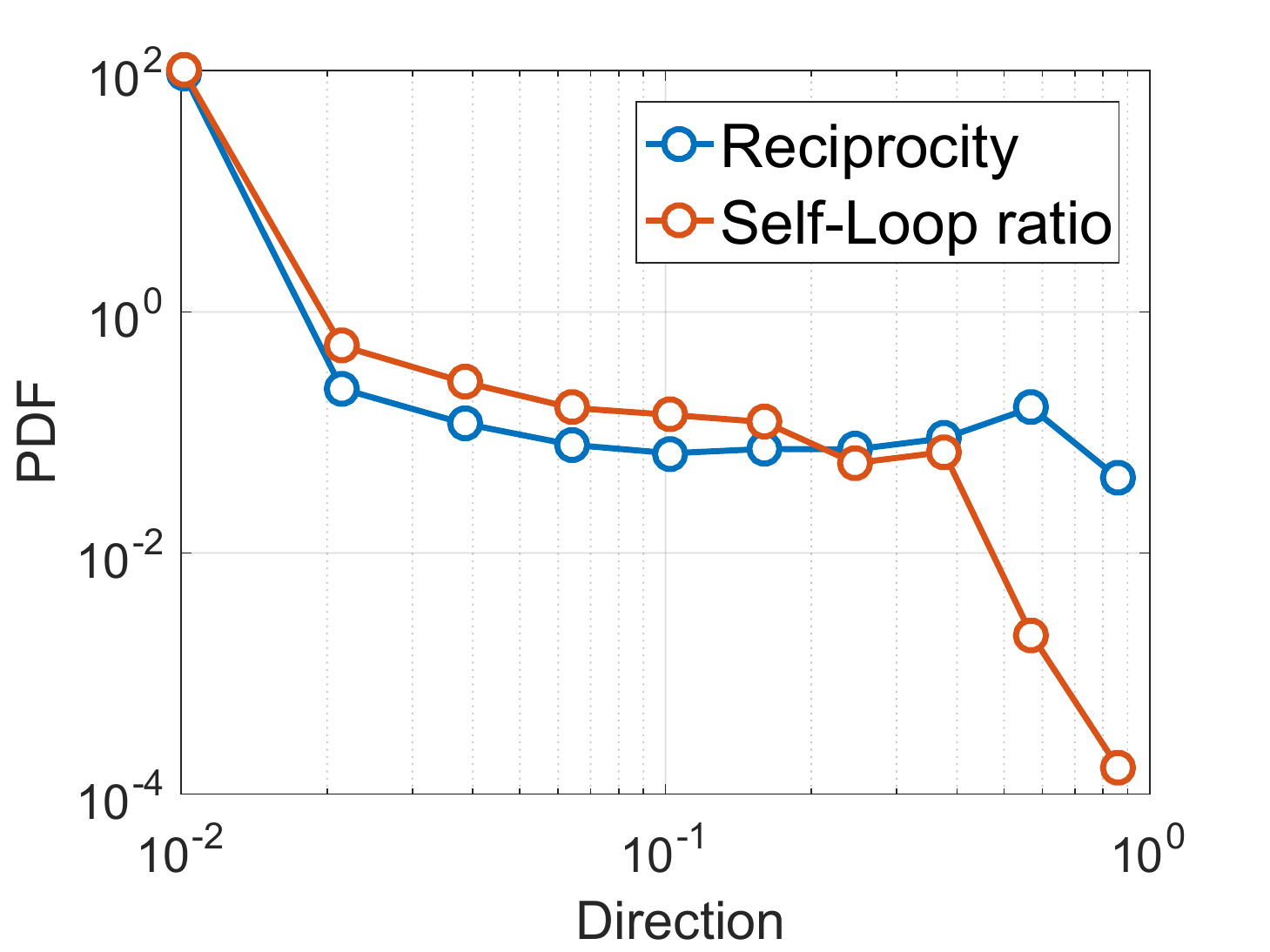}
\label{fig:Intro2}
}
\caption{ \emph{Cascades have four directions: branching-out, converging-in, reciprocal and self-loop.} (a) The Venn diagram of converge, reverse and self-loop structures. (b) The distributions of branch deviation and converge deviation values. (c) The distributions of the number of reciprocal edges and the number of self-loops. (d) The distributions of the reciprocity value and the self-loop ratio.
\label{fig:texture1d}}
\vspace{-0.2in}
\end{figure} 
We then try to characterize the directions of information flow within a cascade.
The direction metrics quantify the preferred direction of information flow in cascade $C$:
\begin{compactitem}
\item \emph{Branch deviation} measures to what extent the edges in cascade $C$ spreading out to different nodes, characterized by the coefficient of variation of out-degree distribution $p(k_{out})$ of $C$, namely $\frac{std(p(k_{out}))}{mean(p(k_{out}))}$, and $k_{out}(u) = \sum_{v}{\Indicator{\{(u,v)\in E\}}}$ is the out-degree of node $u$ and $\Indicator$ is the indicator function. A large branching coefficient value of $C$ means the edges in $C$ spread out from few nodes to a large amount of nodes, implying the directions of information flow are fully random rather than spreading along a preferred direction.

\item \emph{Converge deviation} measures to what extent the edges in cascade $C$ converging into one node, characterized by the coefficient of variance of in-degree distribution $p(k_{in})$ of $C$, where $k_{in}(v) = \sum_{u}{\Indicator{\{(u,v)\in E \ \& \ u \neq v\}}}$ is the in-degree of node $v$. A cascade with large converging coefficient value indicates a large proportion of edges pointing to few nodes, implying that the information flows into few users.

\item \emph{Reciprocity} measures to what extent the edges in cascade  $C$ pointing to the  reciprocal direction, characterized by the ratio of the number of reciprocal edges to the total number of edges, i.e., $\frac{|\{(u,v)|(u,v) \in E \ \& \  (v,u) \in E \ \& \ u \neq v\}|}{|E|} $. For instance, reciprocity $1$ of $C$ indicates that for each edge $(u,v)$ there exists reciprocal edge $(v,u)$, implying that  information flow spreads from user $u$ to user $v$ and vise versa.

\item \emph{Self-loop ratio} measures to what extent the edges in $C$ starting and pointing to the same direction, characterized by the ratio of the number of nodes which have self-loop edge to the total number of nodes, i.e., $\frac{|\{u|(u,u) \in E \}|}{|V|} $. For instance, self-loop ratio $1$ of $C$ indicates that each node in $C$ has self-loop edge , implying that user $u$ retweets himself and information flow does not spread away.
\end{compactitem}

\mytag{Cascades have four directions.}
Existing studies of cascade focus mainly on the branching-out direction of information flows, but we find  other three directions are also ubiquitous. Specifically, cascades which have at least one of the converge, reverse or self-loop structures, as shown in Fig.~\ref{fig:texture1d}a, accounting for $20.0\%$ of the total population. In addition, cascades usually show a combination of different spreading directions. The Venn diagram in Fig.~\ref{fig:texture1d}a shows the number of cascades with different direction types and their logical relationships. In total, $9.2\%$ of the total cascades have more than two texture types, as shown in the overlapping region of Fig.~\ref{fig:texture1d}. 
The branch coefficient distribution shows a bimodal distribution where two modes are near zero and two respectively, implying that information flow in cascades tend to spread along one direction, or a moderate number of directions.  In addition, very large values of branch coefficient do exist. 

 \mytag{Bimodal law for reciprocal edges, and power law for the self-loop edges.} 
As shown in Fig.~\ref{fig:texture1d}c, we find the number of reciprocal edge for each cascade follows the bimodal law as indicated by Equation~\ref{equ:bimodal}. Table 1 shows the best fitting results. The bimodal distribution of the reciprocal edges will be further illustrated by the joint distribution in Fig.~\ref{fig:allVSmass} e. 
We also fit the self-loop edge count by Equation~\ref{equ:bimodal} with $c2=0$, indicating power law distribution with slope $2.47$. The fat-tailed nature of reciprocal edge and self-loop edges shows the prevalence of these two directions. Further, we plot the distributions of converge deviation, reciprocity and self-loop ratio in Figs.~\ref{fig:texture1d}b\&d, which exhibit a stair-like distribution, namely mode near zero, a flat-like distribution at a moderate value range, and also followed by a fat-tail rage at the large values. 

\subsubsection{Cascade activity}
\begin{figure}[!htb]
\vspace{-0.1in}
\centering
\subfigure[]{
\includegraphics[width=0.3\textwidth, trim = 0 0 0 0, clip]{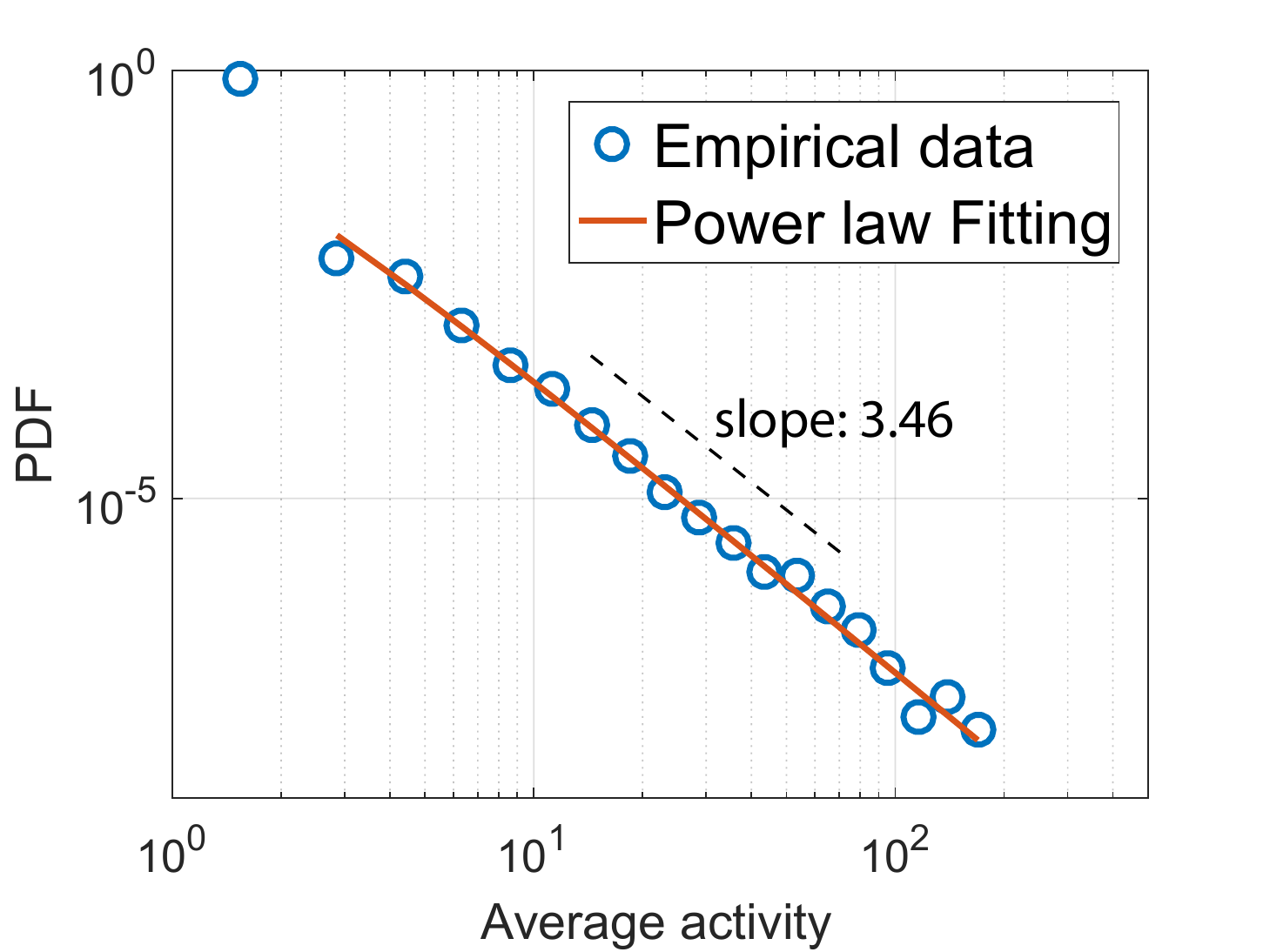}
\label{fig:venn}
}
\vspace{-0.1in}
\caption{ {Average activity follows power law distribution.}
\label{fig:Activity}}
\vspace{-0.1in}
\end{figure} 
Furthermore, a cascade $C$ is a weighted graph due to the fact that a user may retweet other users multiple times in $C$.  We try to quantify multiple user activities in $C$. We define $w(v)=\sum_{u}{w(u,v)}$ as the weight of node $v$, showing that $v$ may retweet more than one user $w(v)$ times in the cascade $C$. Such repeated retweeting activities in $C$ bring large discrepancy between the number of posts and the number of involved users (Fig.~\ref{fig:Activity}  and Fig.~\ref{fig:allVSmass} g). We quantify the user activity of a cascade as follow:
\begin{compactitem}
 \item \emph{Average activity } of cascade $C$ measures extend to which user actively participates in $C$, characterized by the ratio of the number of posts in $C$ to the number of unique users $C$, i.e., $\frac{\sum_{v}{w(v)}}{|V|}$.
\end{compactitem}

\mytag{Average activity power law.}
We find the repeated retweets are prevalent. Cascades with repeated behaviors account for $28.7\%$ of the total population. Furthermore, a seemingly popular cascades with a lot of retweets can collapse to few infected users. We measure the average activity value of each empirical cascade and plot its probability density function in Figure~\ref{fig:Activity}.  We find that the average activity value of each cascade, apart from the value $1$, follows a power law distribution $f(x) = c_1 (x + x_0)^{-\alpha} $ with slope $3.46$ (details in Table 1), implying the existence of very large average activity values. For example, the largest average activity value $165$ belongs to a cascades with $1,535$ users who make $252,878$ retweets.

\section{High-order structure patterns}
The understanding of cascade structure patterns in high-oder metric space are of vital importance to the cascade prediction, clustering, and outlier detection. However, a paucity of previous works did.
Here, we investigate the high-order structural patterns of information cascades by answering four questions as follow: 
\begin{compactitem}
 \item Are cascades wide and shallow? Or narrow and deep? (Sec. 4.1)
 
 \item To what extent do cascades follow the star-like pattern or chain-like pattern? (Sec. 4.2)
 
 \item To what extent do the four directions of information flow coexist in a cascade? (Sec. 4.3)
 
 \item What are the structure patterns of so-called popular cascades? (Sec. 4.4)
\end{compactitem}

\subsection{Size correlations}
\begin{figure}[!htb]
\centering
\subfigure[Mass-Breadth]{
\includegraphics[width=0.145\textwidth, trim = 0 0 5 0, clip]{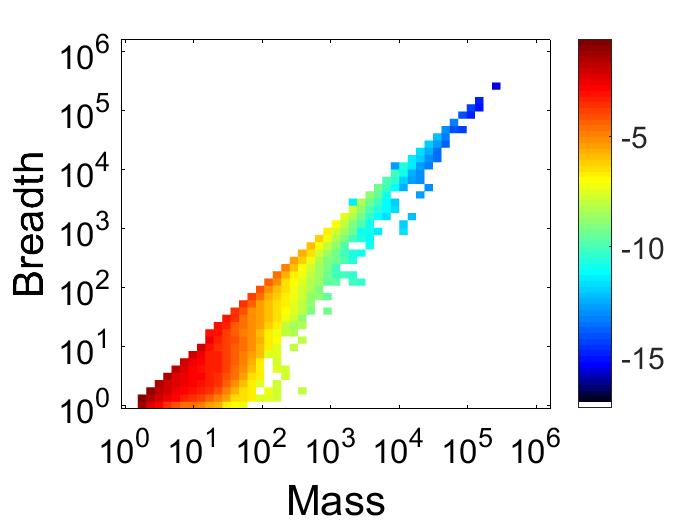}
\label{fig:Intro1}
}
\subfigure[Mass-Length]{
\includegraphics[width=0.145\textwidth, trim = 0 0 5 0, clip]{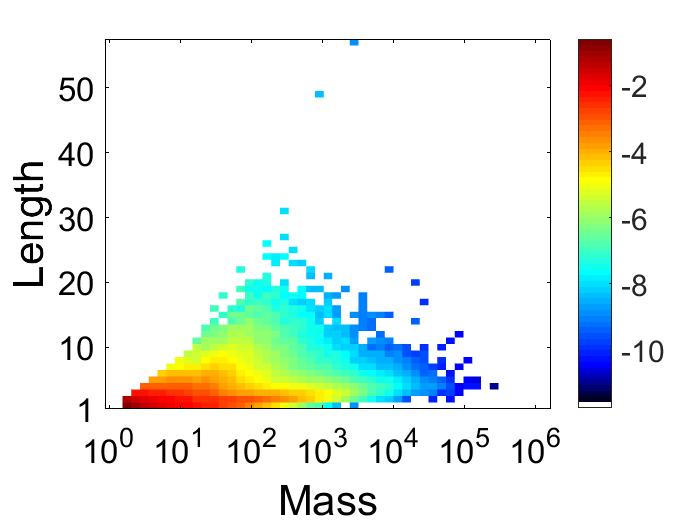}
\label{fig:Intro2}
}
\subfigure[Length-Breadth]{
\includegraphics[width=0.145\textwidth, trim = 0 0 5 0, clip]{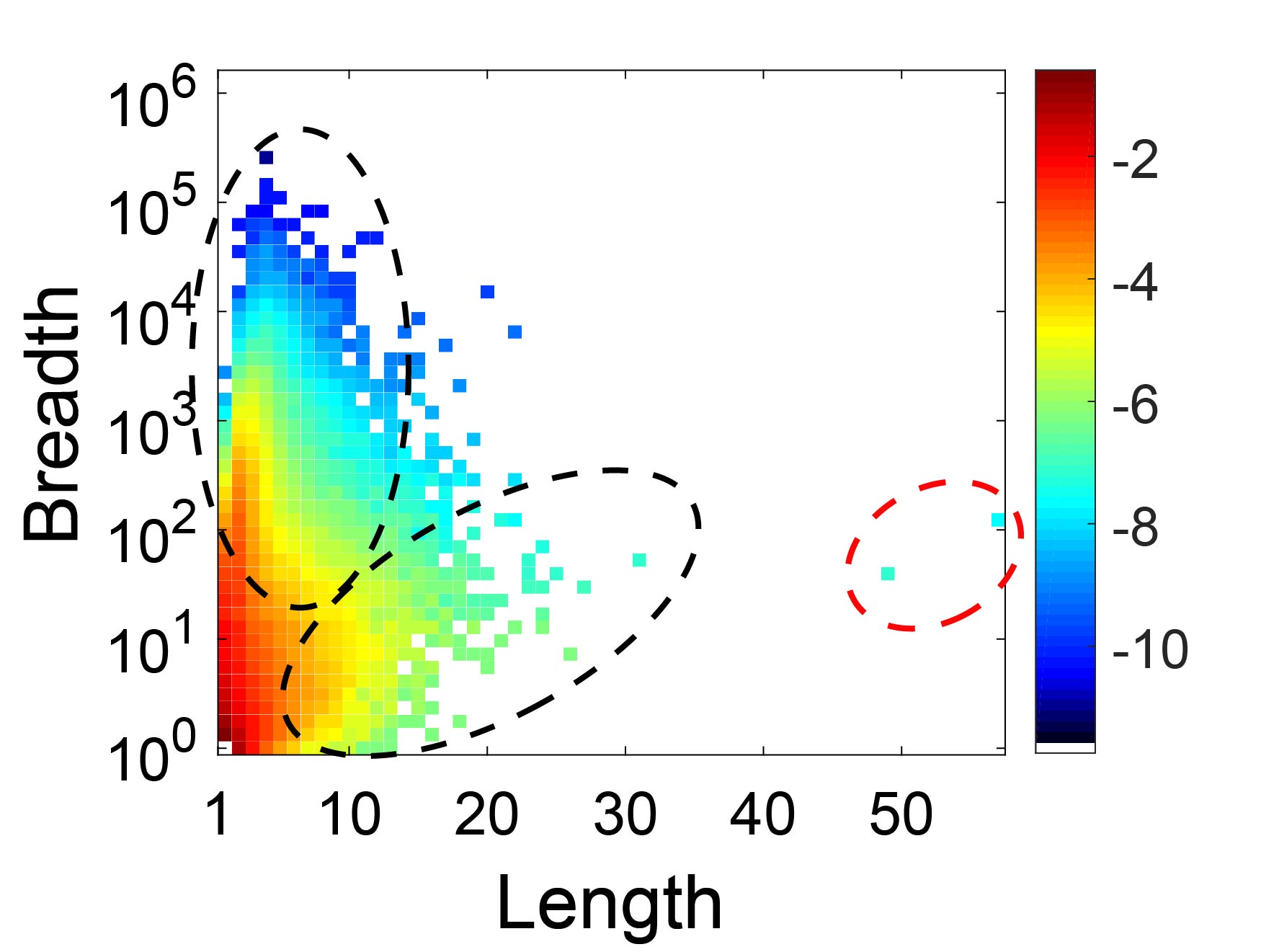}
\label{fig:2dSize}
}
\caption{High-order structure patterns in cascade size metric space. 
\label{fig:2dsize}}
\vspace{-0.1in}
\end{figure} 
\mytag{Breadth dominates mass.}
Figures~\ref{fig:2dsize}a-c plot the joint density profiles for the size metrics. We find large positive correlation between mass and breadth as shown in Fig.~\ref{fig:2dsize}a. Indeed, the correlation coefficient between the logarithmically transformed mass and breadth values is a strikingly high $0.99$, indicating that the breadth accounts for a large proportion of mass, implying the dominant position of the fanning out patterns. 

\mytag{Big, wide and shallow. Or small, narrow and deep.}
Figure~\ref{fig:2dsize}b plots the joint distribution of length and mass, and we observe the biggest cascades are constrained to a relative small length, and the longest cascades are with moderate mass values.  Figure~\ref{fig:2dsize}c plots the joint distribution of length and breadth. We observe most of the cascades are wide and shallow (illustrated in the upper ellipse in Fig.~\ref{fig:2dsize}c). In addition, there indeed exist narrow and deep cascades (illustrated in lower ellipse in Fig.~\ref{fig:2dsize}c). In contrast, it is difficult to find the very wide and deep cascades, or the very narrow and deep cascades. 
The structure patterns of big cascades will further discussed in Section 4.4.

\subsection{Silhouette correlations} 
\begin{figure}[!htb]
\centering
\subfigure[Fluct-Trend]{
\includegraphics[width=0.145\textwidth, trim = 0 0 0 0, clip]{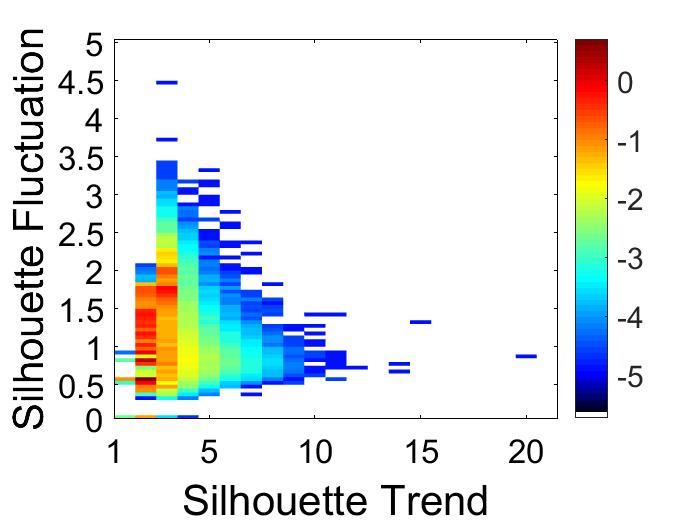}
\label{fig:Intro2}
}
\subfigure[Trend-Length]{
\includegraphics[width=0.145\textwidth, trim = 0 0 0 0, clip]{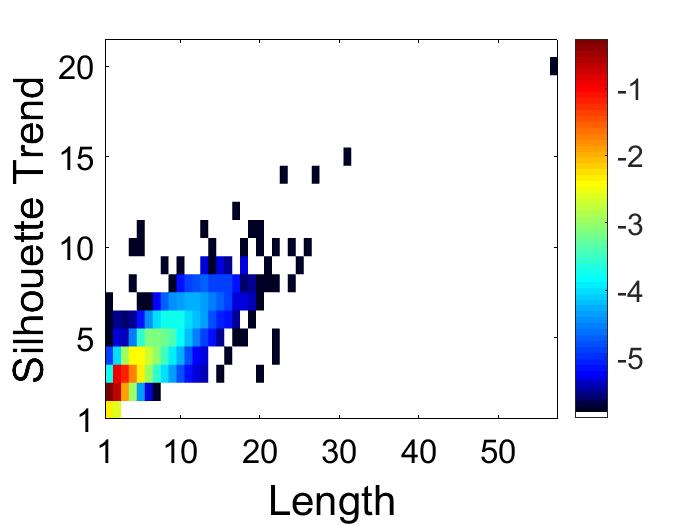}
\label{fig:Intro1}
}
\subfigure[Fluct-Length]{
\includegraphics[width=0.145\textwidth, trim = 0 0 0 0, clip]{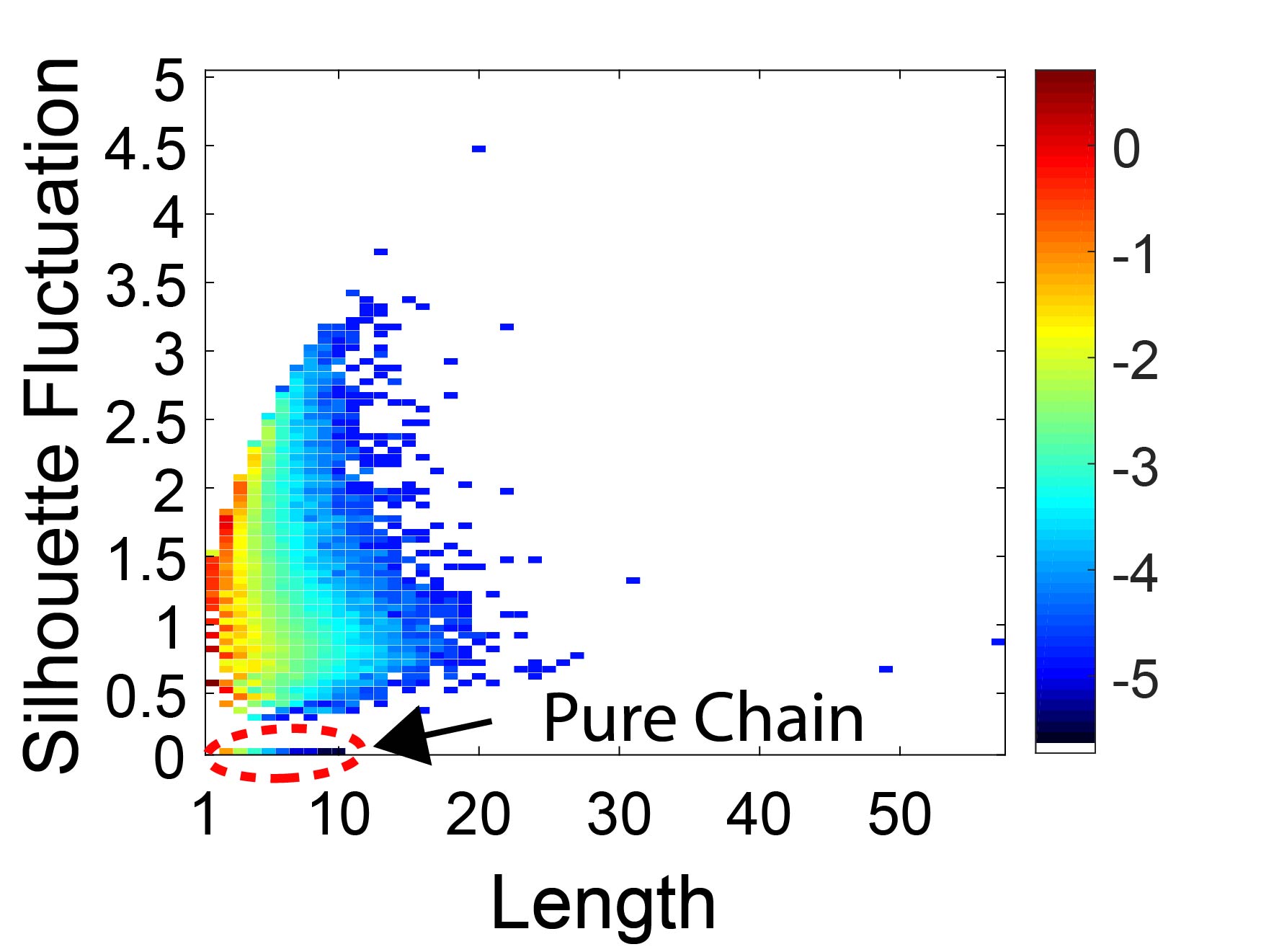}
\label{fig:Intro2}
}
\caption{ High-order structure patterns in cascade silhouette metric space. 
\label{fig:shape}}
\end{figure}

\mytag{The prevalence of star pattern.}
The majority of cascades have silhouette $S_C=1\times B_1$ (star pattern) with trend values $\le 2$, as shown in Figs.~\ref{fig:shape}a\&c, which account for $54.49\%$ 
of the total cascades population ($1,600,228$, with trivial single-node and two-nodes cascades being excluded), which support the findings of the prevalence of small star pattern \cite{leskovec2007patterns, goel2012structure}. 

\mytag{Chain patterns with limited length.}
The $S_C=1^L$ (chain with length $L-1$) cascades account for $7.87\%$ ($126,003$) of the total population, featured with fluctuation value $0$ as shown in Figs.~\ref{fig:shape}a\&c. We highlight the pure chain patterns in Fig.~\ref{fig:shape}c. We find the maximum length of pure chain is $12$, i.e., $1^{13}$ pattern, and as the length $L$ increases the number of pure chain decays quickly as indicated by the color,  implying the limited length of chain pattern in empirical data. 

\mytag{Deep patterns with moderate fluctuations.}
As shown in Fig.~\ref{fig:shape}c, as length increases, the deep patterns have a moderate fluctuation values, implying that the deep pattern relying on the continuously spreading to a moderate users at each depth, rather than on the chain-like information pathway .

\mytag{The attached peculiar long chain.}
We find trend value (the Wiener index) and length are positive correlated in Fig.~\ref{fig:shape}b. However, there still exists variance along the shape trend axis, implying two ways to reach a long length: a star with a peculiar long chain or  viral structures in Ref.\cite{liben2008tracing}. Indeed, the peculiar long chain may increase the fluctuation values and length while keep relatively stable trend values (red regions in  Figs.~\ref{fig:shape}a\&c).


\subsection{Direction correlations} 
\begin{figure}[!htb]
\centering
\subfigure[Branch-Conv]{
\includegraphics[width=0.145\textwidth, trim = 0 0 0 0, clip]{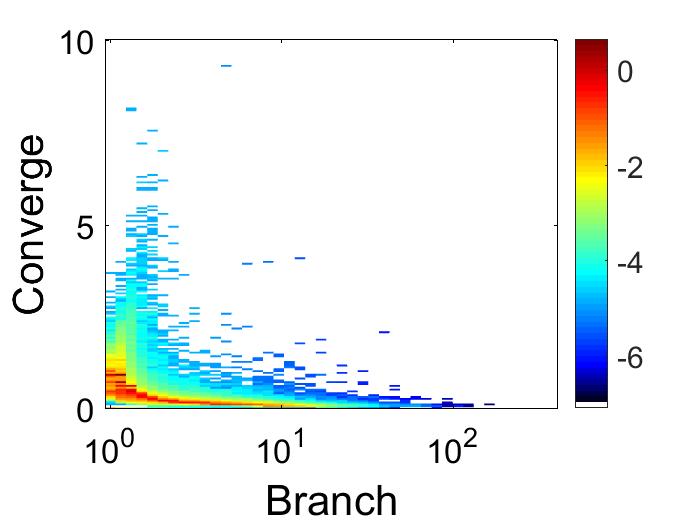}
\label{fig:Intro1}
}
\subfigure[Branch-Recipro]{
\includegraphics[width=0.145\textwidth, trim = 0 0 0 0, clip]{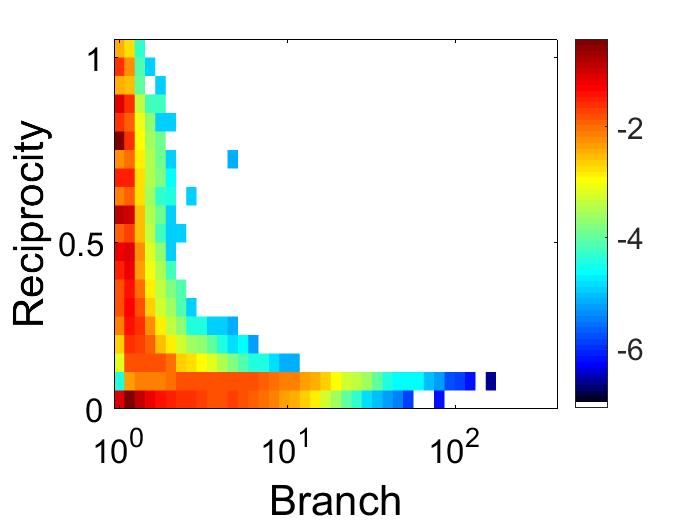}
\label{fig:Intro2}
}
\subfigure[Branch-Loop]{
\includegraphics[width=0.145\textwidth, trim = 0 0 0 0, clip]{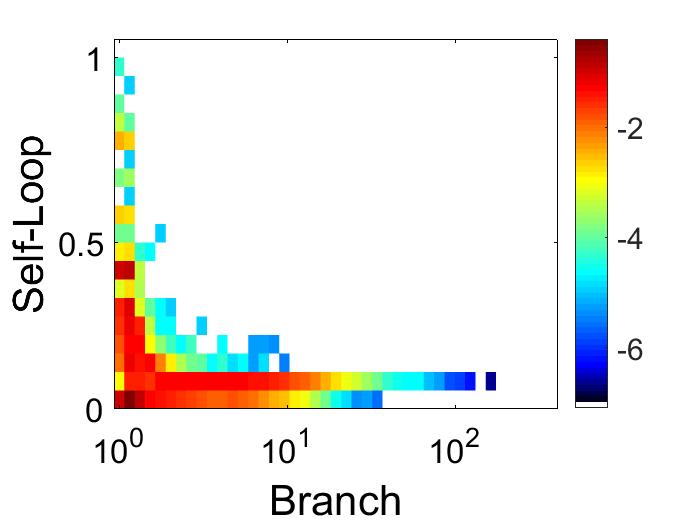}
\label{fig:Intro2}
}
\subfigure[Conv-Reverse]{
\includegraphics[width=0.145\textwidth, trim = 0 0 0 0, clip]{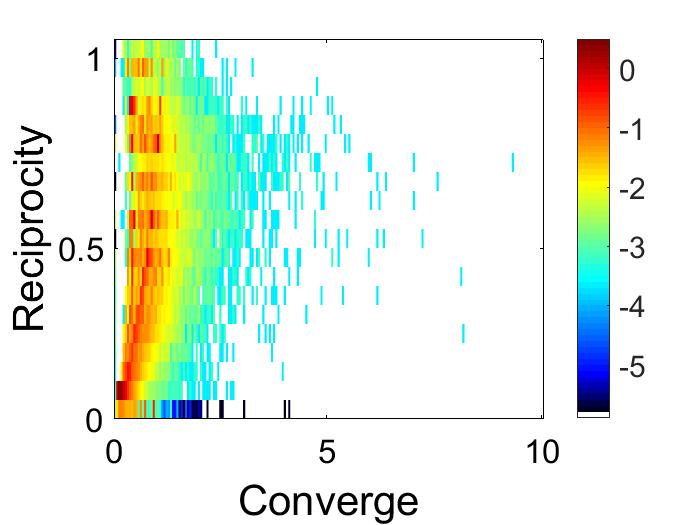}
\label{fig:Intro1}
}
\subfigure[Conv-Loop]{
\includegraphics[width=0.145\textwidth, trim = 0 0 0 0, clip]{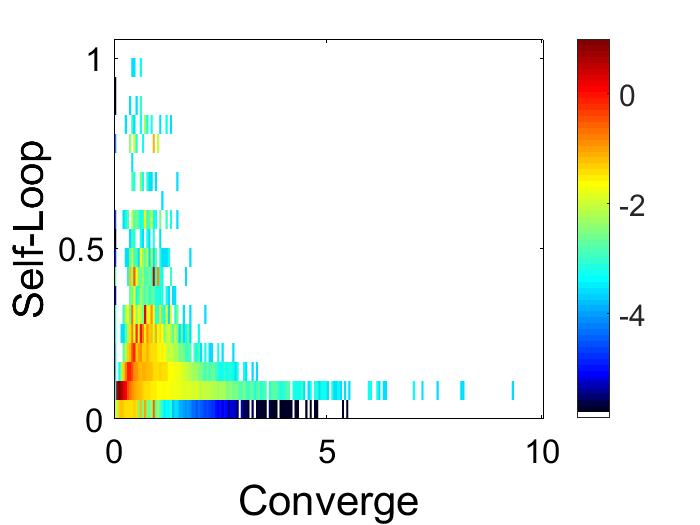}
\label{fig:Intro2}
}
\subfigure[Recipro-Loop]{
\includegraphics[width=0.145\textwidth, trim = 0 0 0 0, clip]{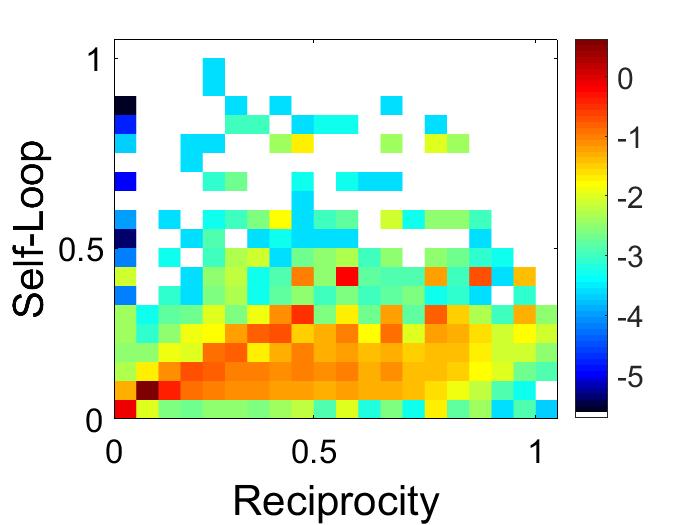}
\label{fig:Intro2}
}
\caption{ High-order structure patterns in cascade direction metric space.
\label{fig:texture}}
\end{figure} 
Information spreadings follow four direction as discussed in above, i.e., branching-out, converging-in, reciprocal and self-loop. Here, we examine to what extent that these four spreading directions can coexist. 

\mytag{``L''-shape non-coexistence relationships.}
We find non-coexistence relationships, featuring ``L'' shape in joint distribution, between these direction metrics. The branch direction and converge direction show non-coexistence relationship just like two polarities. For instance, Figure~\ref{fig:texture}a plots the heat map of branch deviation vs. converge deviation value for each cascade. We find that large converge values only exist with small branch values, and vice versa. These kind of  non-coexistence  relationships are also applied to branch vs. reciprocal (Fig.~\ref{fig:texture}b), branch vs. self-loop (Fig.~\ref{fig:texture}c), and converge vs. self-loop (Fig.~\ref{fig:texture}e).

\mytag{Coexistence relationships.}
  In contrast, reciprocity shows little correlation between converge deviation value (Fig.~\ref{fig:texture}d) or self-loop ratio(Fig.~\ref{fig:texture}f), implying that either reciprocal vs. converge or reverse vs. self-loop can coexist in real information spreadings.

\subsection{Structural patterns of big cascades}

To predict the so-called popular or big cascades is of vital importance. However, without understandings of the structure patterns of big cascades, modeling or predicting the spreading process in social media remains a challenge.
Here, we investigate the high-order structure patterns of the big cascades.

\begin{figure}[!t]
\vspace{-0.1in}
\centering
\subfigure[Trend-Mass]{
\includegraphics[width=0.145\textwidth, trim = 0 0 0 0, clip]{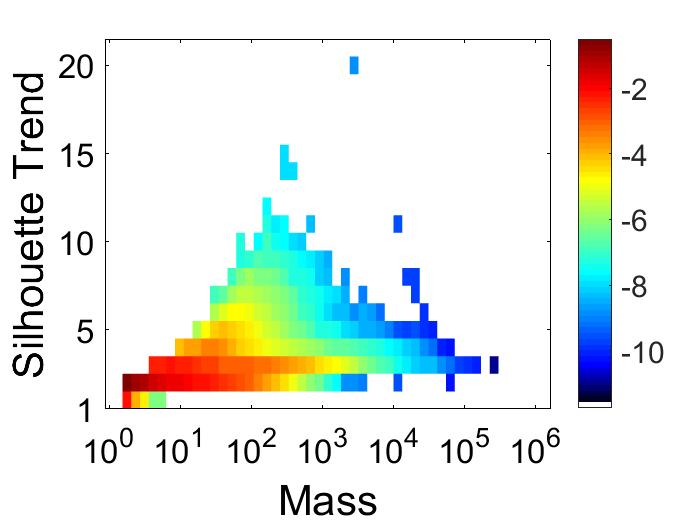}
\label{fig:venn}
}
\subfigure[Fluct-Mass]{
\includegraphics[width=0.145\textwidth, trim = 0 0 0 0, clip]{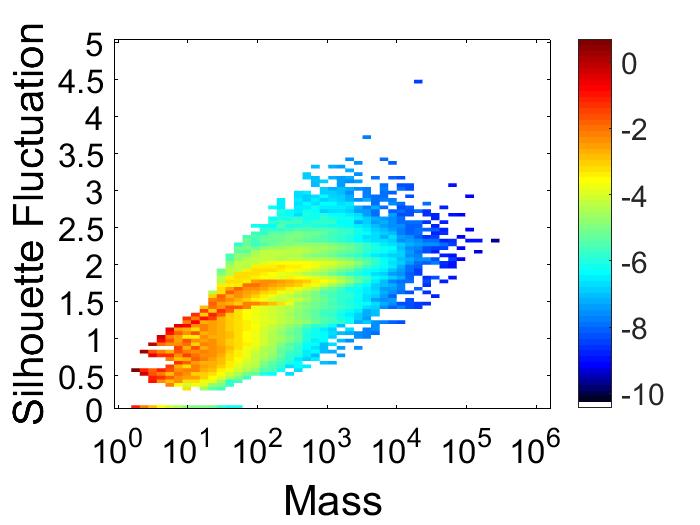}
\label{fig:Intro1}
}
\subfigure[Branch-Mass]{
\includegraphics[width=0.145\textwidth, trim = 0 0 0 0, clip]{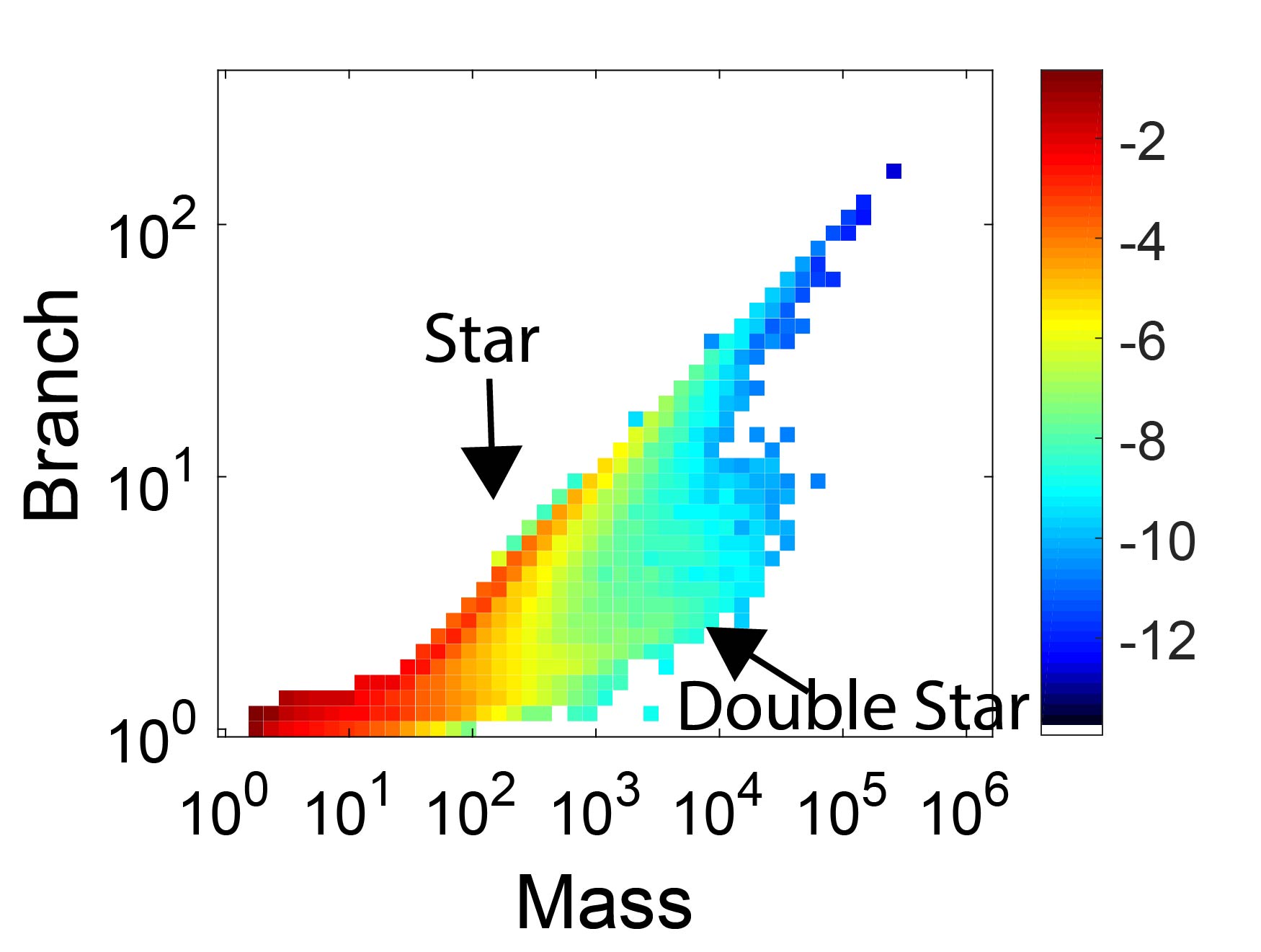}
\label{fig:Intro2}
}
\subfigure[Conv-Mass]{
\includegraphics[width=0.145\textwidth, trim = 0 0 0 0, clip]{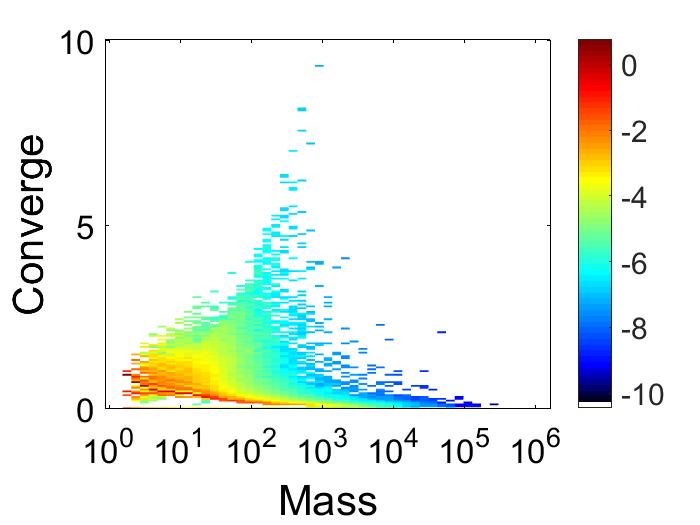}
\label{fig:Intro1}
}
\subfigure[Mass-Reciprocity]{
\includegraphics[width=0.145\textwidth, trim = 0 0 0 0, clip]{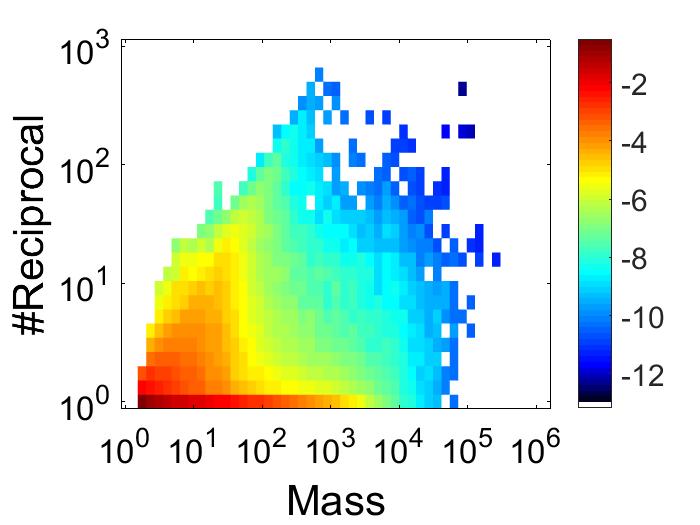}
\label{fig:Intro1}
}
\subfigure[Mass-Loop]{
\includegraphics[width=0.145\textwidth, trim = 0 0 0 0, clip]{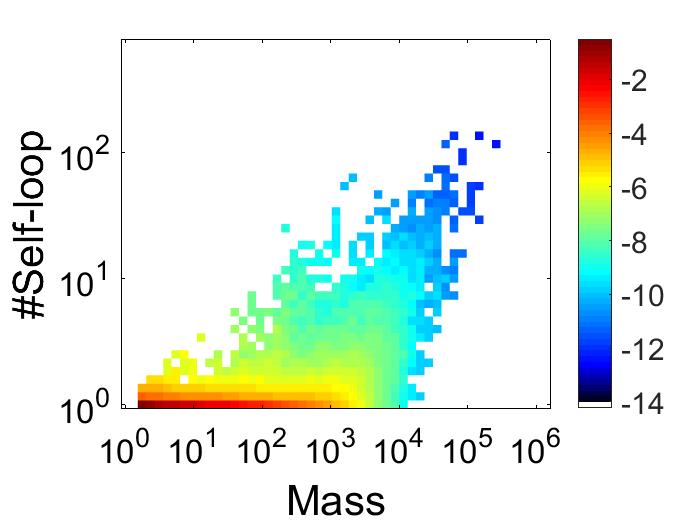}
\label{fig:Intro1}
}
\subfigure[Mass-Retweets]{
\includegraphics[width=0.145\textwidth, trim = 0 0 0 0, clip]{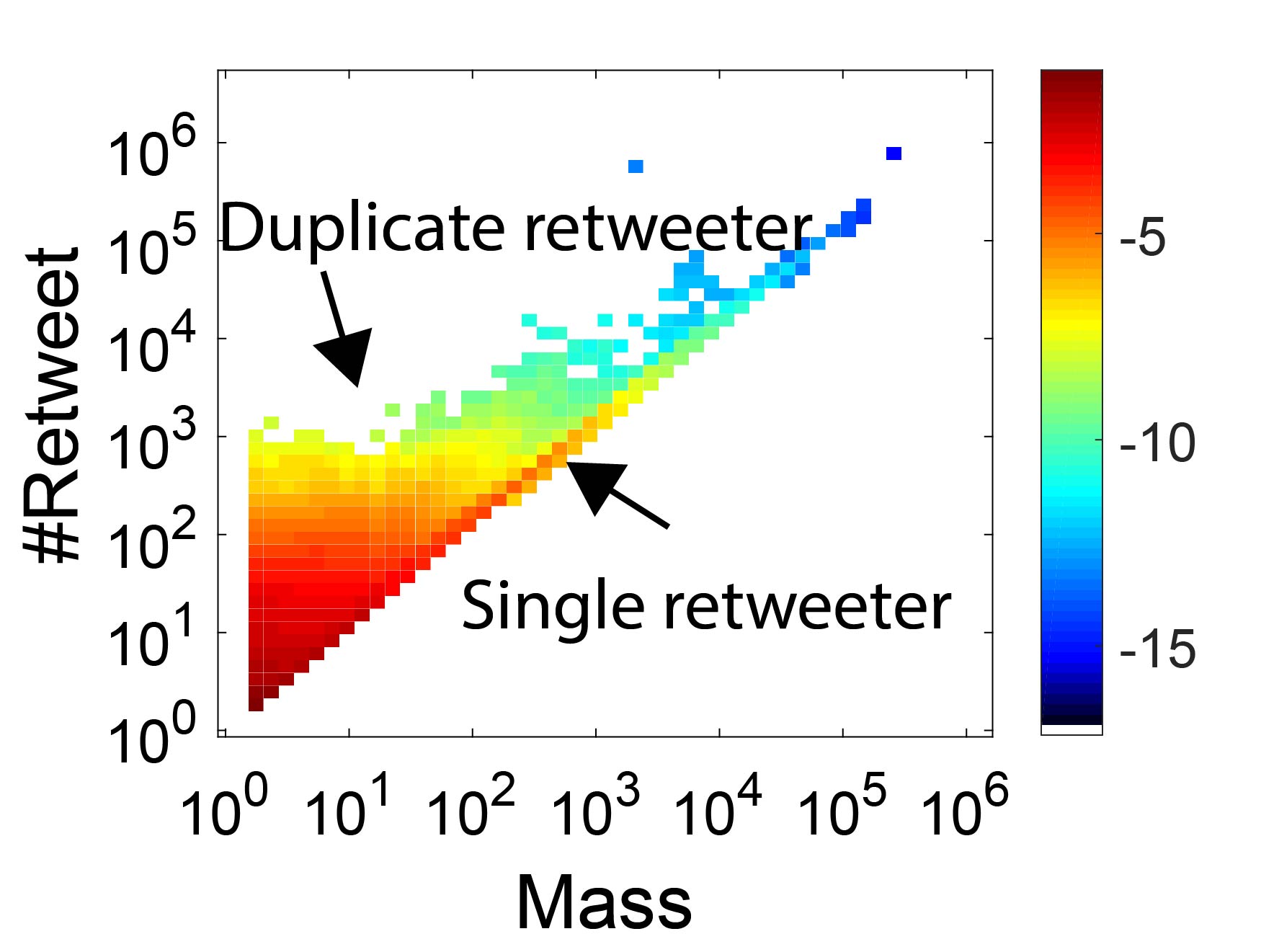}
\label{fig:Intro1}
}
\vspace{-0.1in}
\caption{High-order structure patterns of empirical cascades as mass increases.
\label{fig:allVSmass}}
\end{figure}

\mytag{Big cascades tend to spread along breadth dimension, while medium-sized cascades exhibit rich structure patterns.}
We first try to answer what kinds of silhouette do big cascades have.
Figure~\ref{fig:allVSmass}a plots the joint distribution of silhouette trend and mass. We find cascades tend to follow two trends: spreading along the breadth dimension, or along the depth dimension.
For example, a star graph with silhouette $1\times(N-1)$, and a chain graph with silhouette $1^N$ are two extreme situations of the breadth dominated pattern and the chain dominated pattern respectively. The silhouette trend values (wiener index) of above star and chain graph are $2-\frac{2}{N}$ and $\frac{N+1}{3}$ respectively as shown in the supporting information of Ref. \cite{Zang2017}. Indeed, $2-\frac{2}{N}$ and $\frac{N+1}{3}$ are the floor boundary and ceiling boundary respectively as shown in Fig.~\ref{fig:allVSmass}a.

Furthermore, the empirical data shows that the breadth dominated patterns lie in the region stretching into the largest mass values with  trend values $\le 5$, indicating that the big cascade tend to follow breadth dominated patterns.
We also find large variance of  trend values near moderate mass $[10^2,10^3]$ as shown in Fig.~\ref{fig:allVSmass}a, implying that  moderate cascades can exhibit a wide range of structures from breadth dominated patterns to depth dominated patterns.


\mytag{Big cascades are attached with some peculiar chains.}
Figure~\ref{fig:allVSmass}b plots the joint distribution of silhouette fluctuation and mass of each cascade. The breadth dominated patterns usually exhibit large fluctuation values, while depth dominated patterns show small fluctuation values. For example, a star graph with shape $1\times(N-1)$ has fluctuation value $\sqrt{2}(1-\frac{2}{N})$, and a chain graph with shape $1^N$ has fluctuation value $0$. We further find fluctuation values even capture the minor difference of the star-like graphs $1\times(N-K-1)\times1^K, K=0,1,...$. By calculating the coefficient of variation of sequence $1, N-K-1, 1,...,1$, we find the fluctuation values for star-like graphs $1\times(N-K-1)\times1^K$ taking on $\sqrt{2+K}(1-\frac{2+K}{N})$, as shown a family of red curves in Fig.~\ref{fig:allVSmass}b. 

\mytag{Big cascades may consist of multiple branches of stars, like double-star, but the biggest ones are dominated by one branch of star.}
We then investigate the correlations between cascade directions and its mass. We plot the  joint distributions of branch deviation vs. mass in Fig.~\ref{fig:allVSmass}c.
We find two clusters in Fig.~\ref{fig:allVSmass}c as mass increases, indicating that cascades are dominated by one branch of star pattern (red region in diagonal direction,) or multiple branches of star patterns (cluster below the diagonal). For example, a star graph $1 \times (N-1)$ has branch value $\sqrt{N}$ by calculating the coefficient variation of out-degree distribution of the star graph, where a double-star graph $1 \times (\frac{N}{2}-1) \times \frac{N}{2}$ has branch value $\approx \sqrt{ \frac{N}{2}}$.  However, we find the biggest cascades, in the up-right most region, are dominated by one major star patterns.

\mytag{Big cascades have less converge hubs.}
Figure~\ref{fig:allVSmass}d plots converge deviations vs. mass. We find two obvious converging trends as mass $N$ increases: non-converging cluster with very small converge value which stretches into very large mass, and converging cluster which peaks converge value at mass near $1000$ magnitude. Our common understanding
of cascades are in non-converging trend which means information spreads out. In contrast, we find cascades with very large converge value, implying the information converge into one user. One plausible explanation is that the few users frequently retweet other posts in one cascade to make this cascade more successful. However, we find this kind of manipulation of information flow can not make a cascade very large. Indeed, as shown in Fig.~\ref{fig:allVSmass}d, cascades with large converge value usually have mass between $100$ and $1000$, while very large cascades show very small converge coefficient value. 
Multiple retweets flowing into a user, say $u$, may help in exhausting $u$'s followers, but these kinds of behaviors may not induce further spreads.  

\mytag{Big cascades have some reciprocal edges and self-loops.}
Figure~\ref{fig:allVSmass}e plots the joint distribution of mass and the number of reciprocal edge. We find, as shown in the reddest region, small cascades have a lot of reciprocal edges while as mass grows the number of reciprocal edges are small. However, as for the very large cascade, there also exist many reciprocal edges.  Figure~\ref{fig:allVSmass}f plots the joint distribution of mass and the number of self-loops. We also find big cascades have many self-loops.


\mytag{Big cascades and borrowed prosperity.}
Figure~\ref{fig:allVSmass}g plots the joint distribution of mass and the number of retweets. We find that
relatively small cascades with mass $\le 1000$ have frequent retweet activities, leading to complex structure patterns of cascades, e.g., the converge, reciprocal and self-loop structure, as discussed in above. However, cascades with large average activity values also show borrowed prosperity because under a large number of retweets there are much fewer actually infected users.
For example, some cascades with $\sim 10^3$ number of retweets are generated by quite a few users as shown in Fig.~\ref{fig:allVSmass}g.
In contrast, very big and successful cascades show relatively equal retweet number 
 (Fig.~\ref{fig:allVSmass}g).

\section{Implications for dynamics}
In the temporal dimension, cascades evolve over time, exhibiting rich temporal patterns. However. it remains an open question that how the dynamics of cascades interplay with their structures. In this section, we explore the structure patterns underpinning these temporal patterns.

\subsection{Characterizing cascade dynamics}
Cascades exhibit rich temporal dynamics which are represented by the variations of their growth rates $c(t)$. Thus, we can characterize cascade dynamics based on their variation patterns by clustering methods. 
We first discuss the high level intuitions of clustering methods, and then get the representative dynamic patterns to characterize cascade dynamics.

\begin{figure}[htb]
\centering
\subfigure[C1: Spike]{
\includegraphics[width=0.147\textwidth, trim = 0 0 0 0, clip]{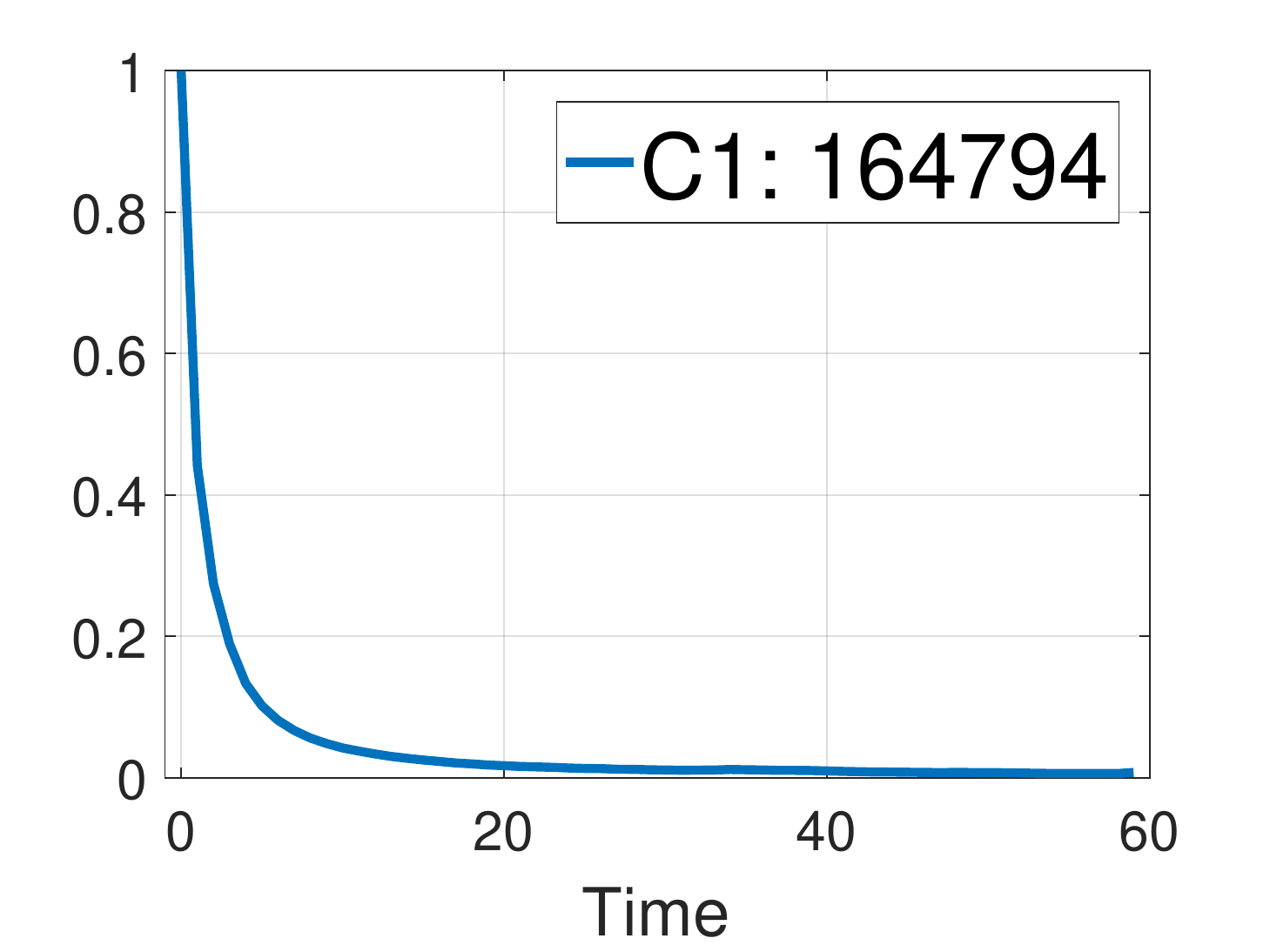}
\label{fig:venn}
}
\subfigure[C2: Fat-Spike]{
\includegraphics[width=0.147\textwidth, trim = 0 0 0 0, clip]{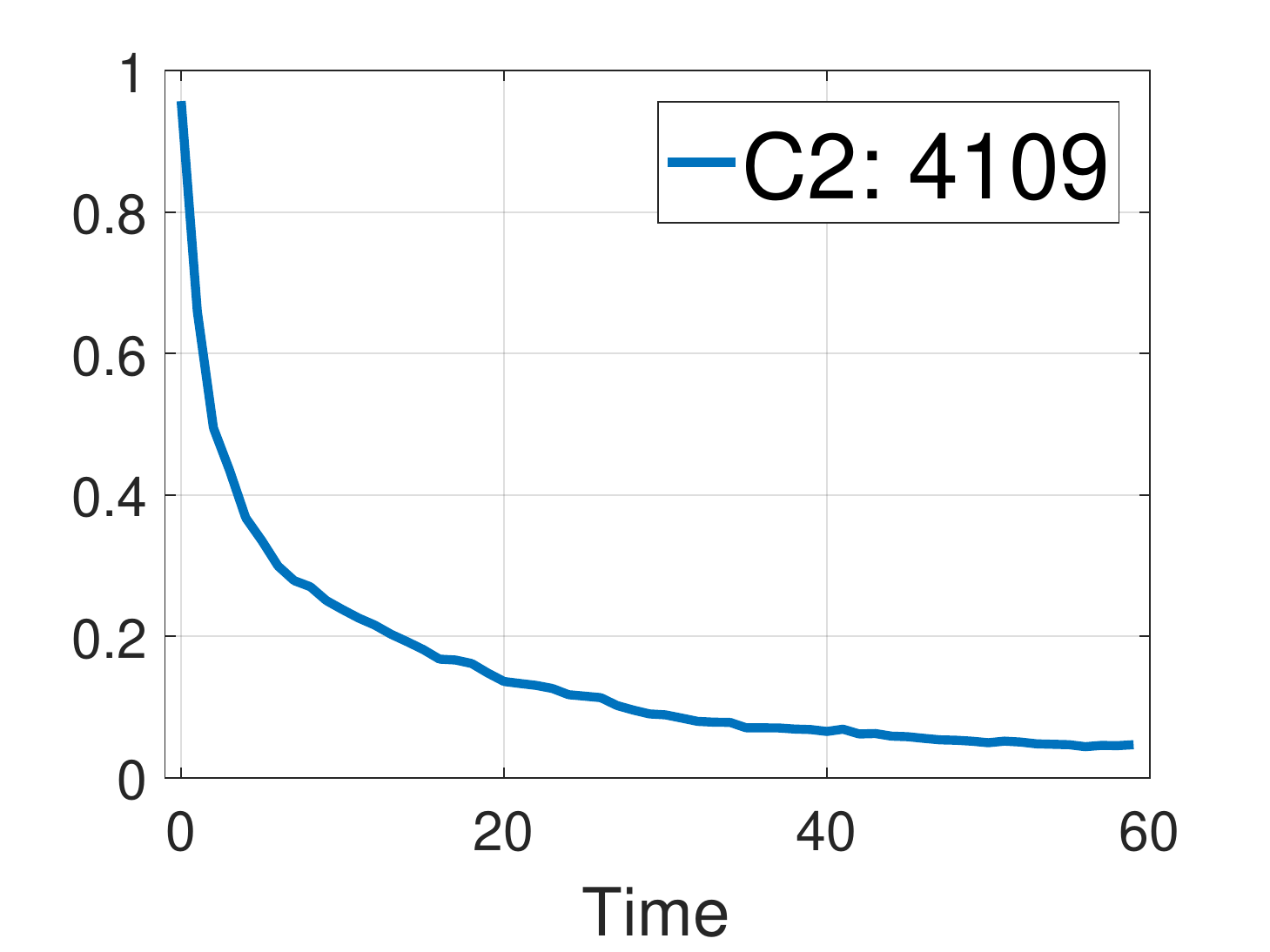}
\label{fig:Intro1}
}
\subfigure[C3: Fatter-Spike]{
\includegraphics[width=0.147\textwidth, trim = 0 0 0 0, clip]{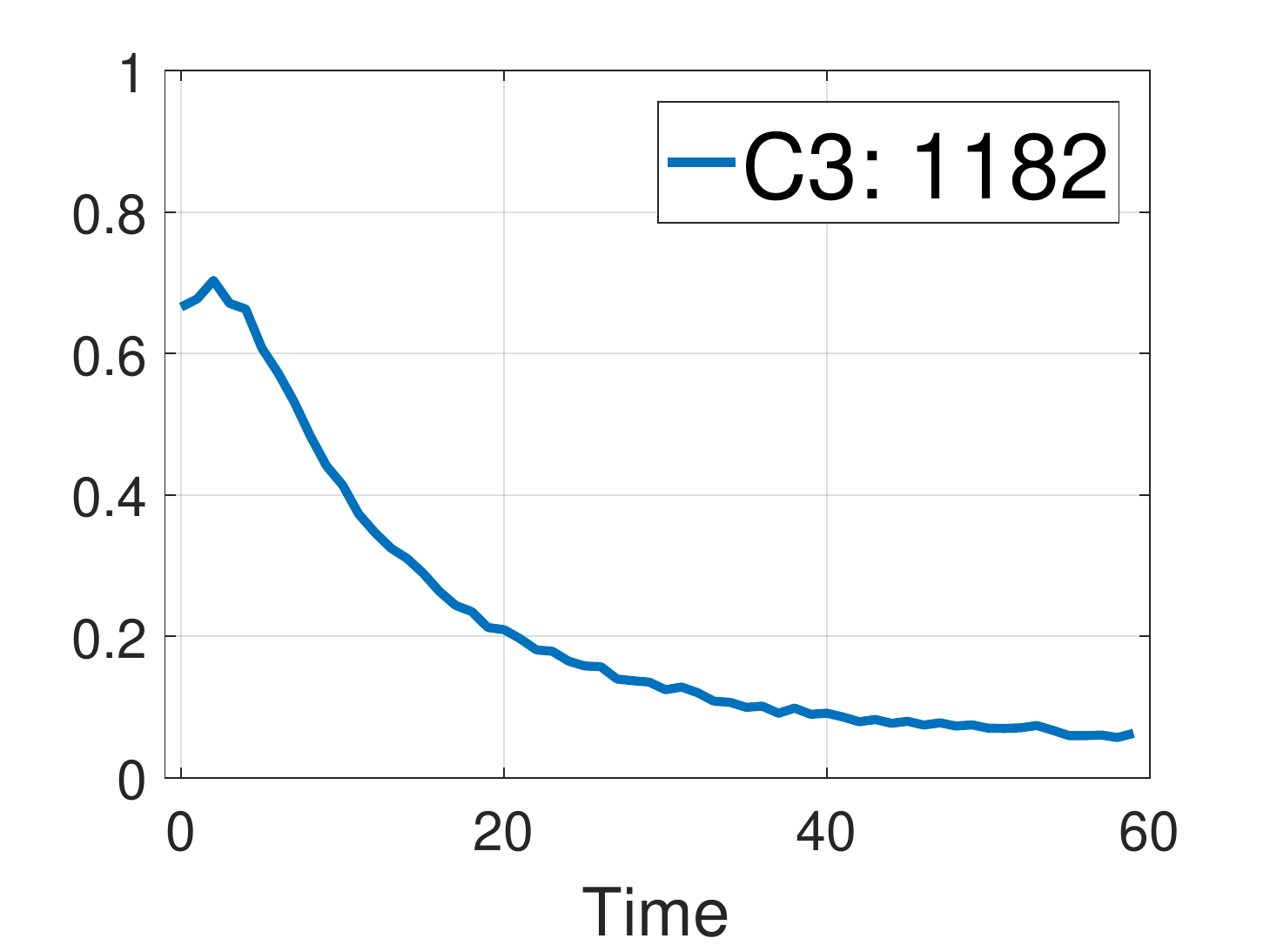}
\label{fig:Intro2}
}
\subfigure[C4: Small-Rebnd]{
\includegraphics[width=0.147\textwidth, trim = 0 0 0 0, clip]{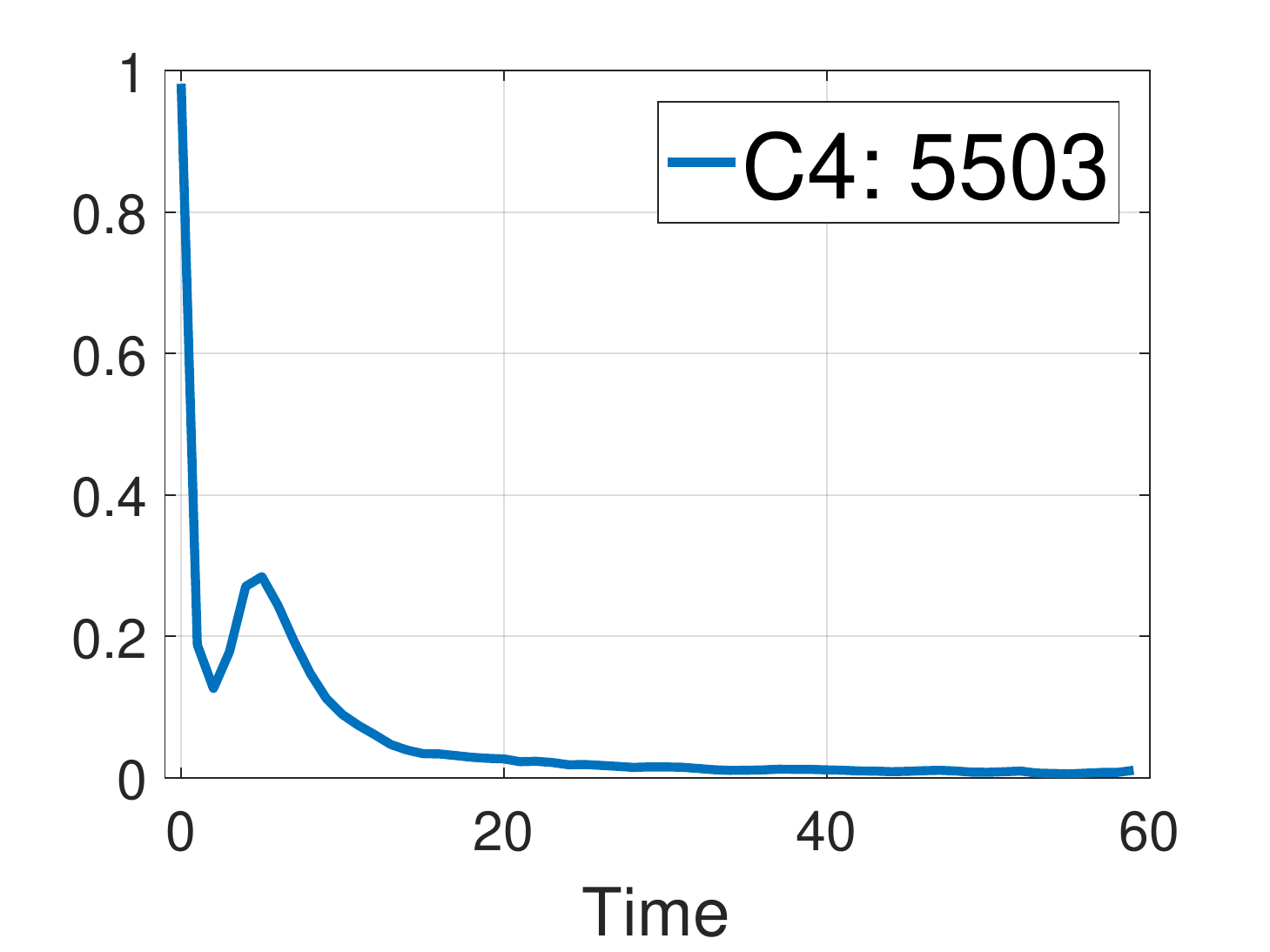}
\label{fig:Intro2}
}
\subfigure[C5: Big-Rebnd]{
\includegraphics[width=0.147\textwidth, trim = 0 0 0 0, clip]{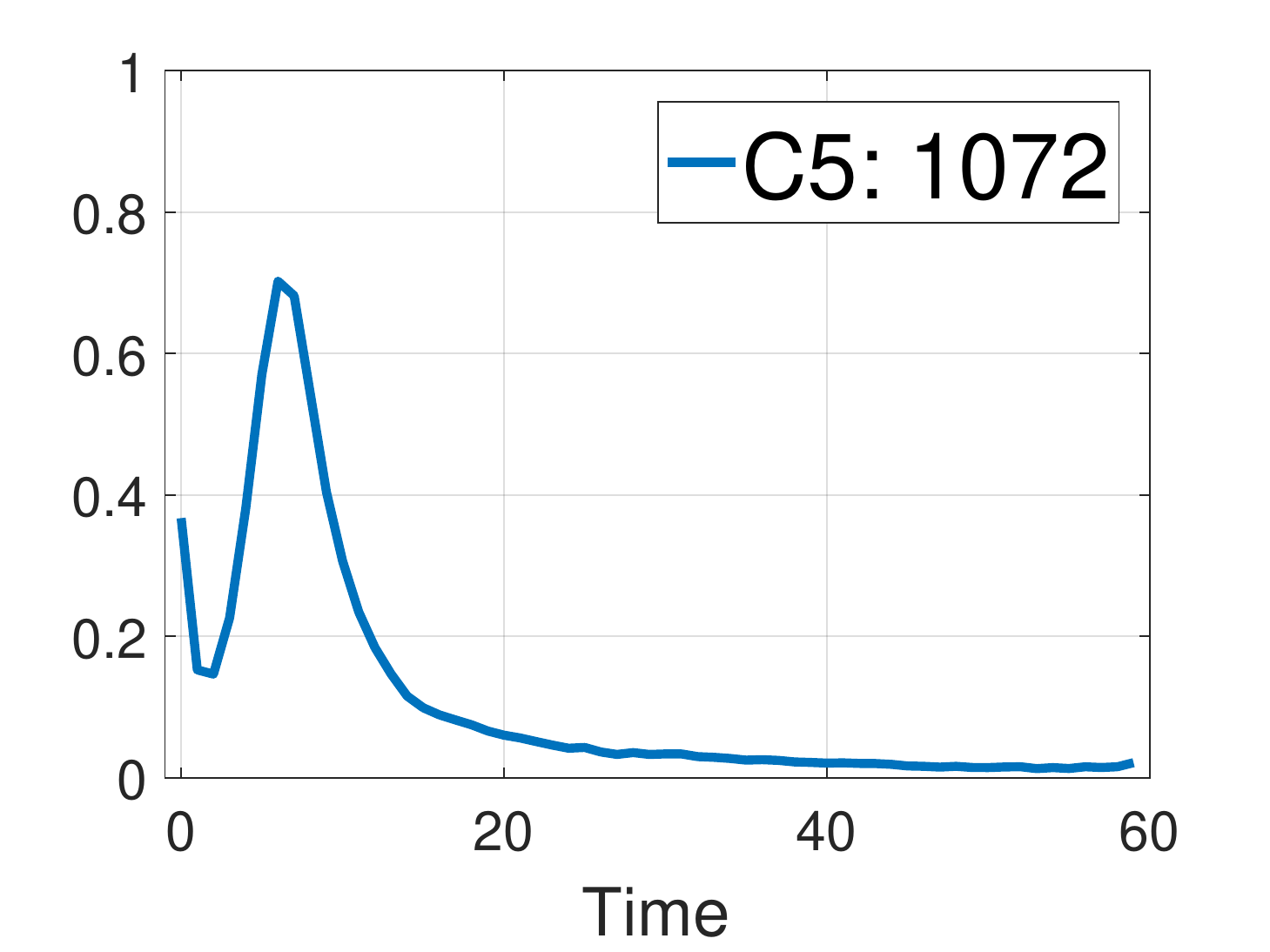}
\label{fig:Intro2}
}
\subfigure[C6: Late-Rebnd]{
\includegraphics[width=0.147\textwidth, trim = 0 0 0 0, clip]{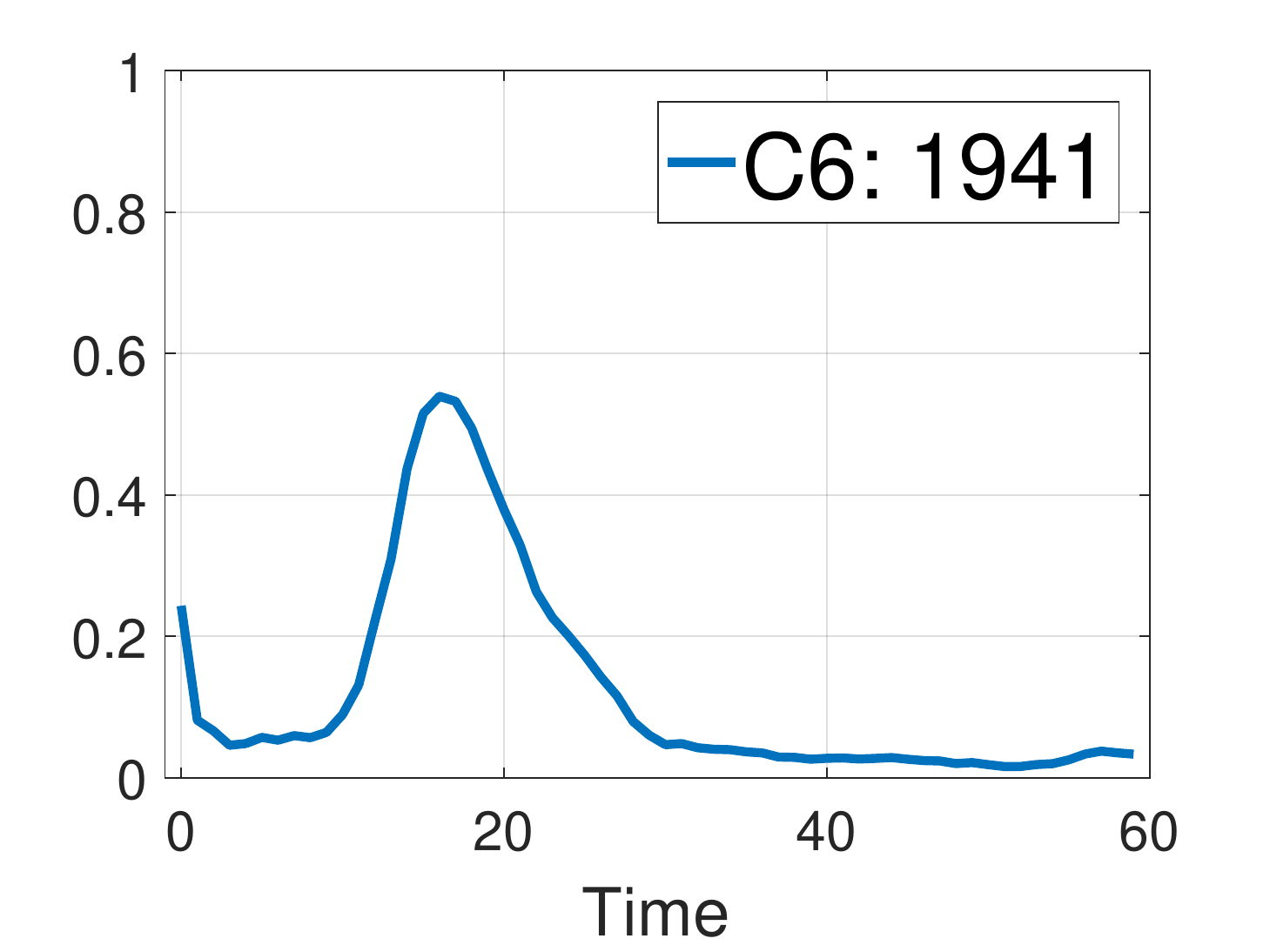}
\label{fig:Intro1}
}
\subfigure[C7: Early-Persist]{
\includegraphics[width=0.147\textwidth, trim = 0 0 0 0, clip]{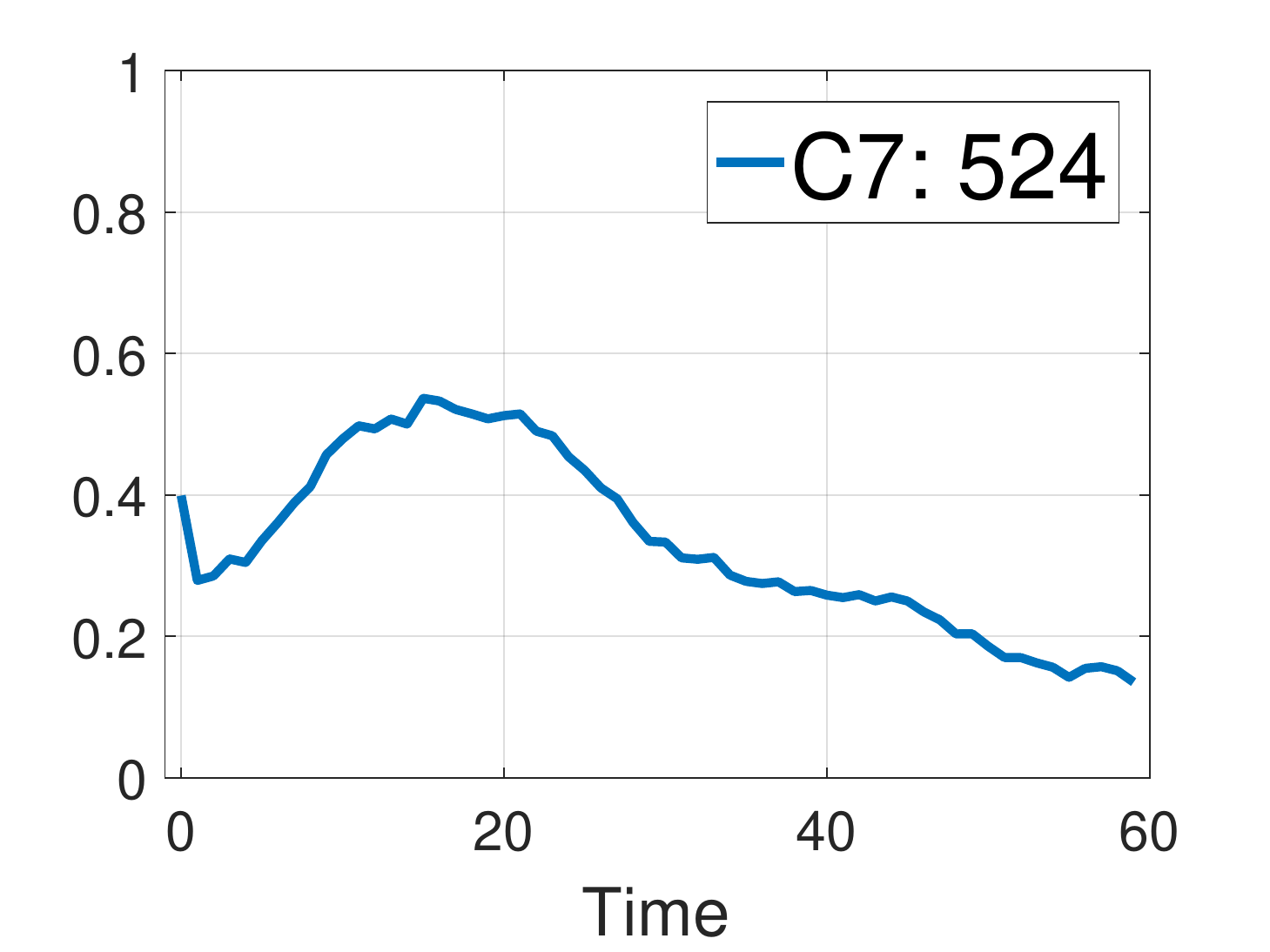}
\label{fig:Intro2}
}
\subfigure[C8: Late-Persist]{
\includegraphics[width=0.147\textwidth, trim = 0 0 0 0, clip]{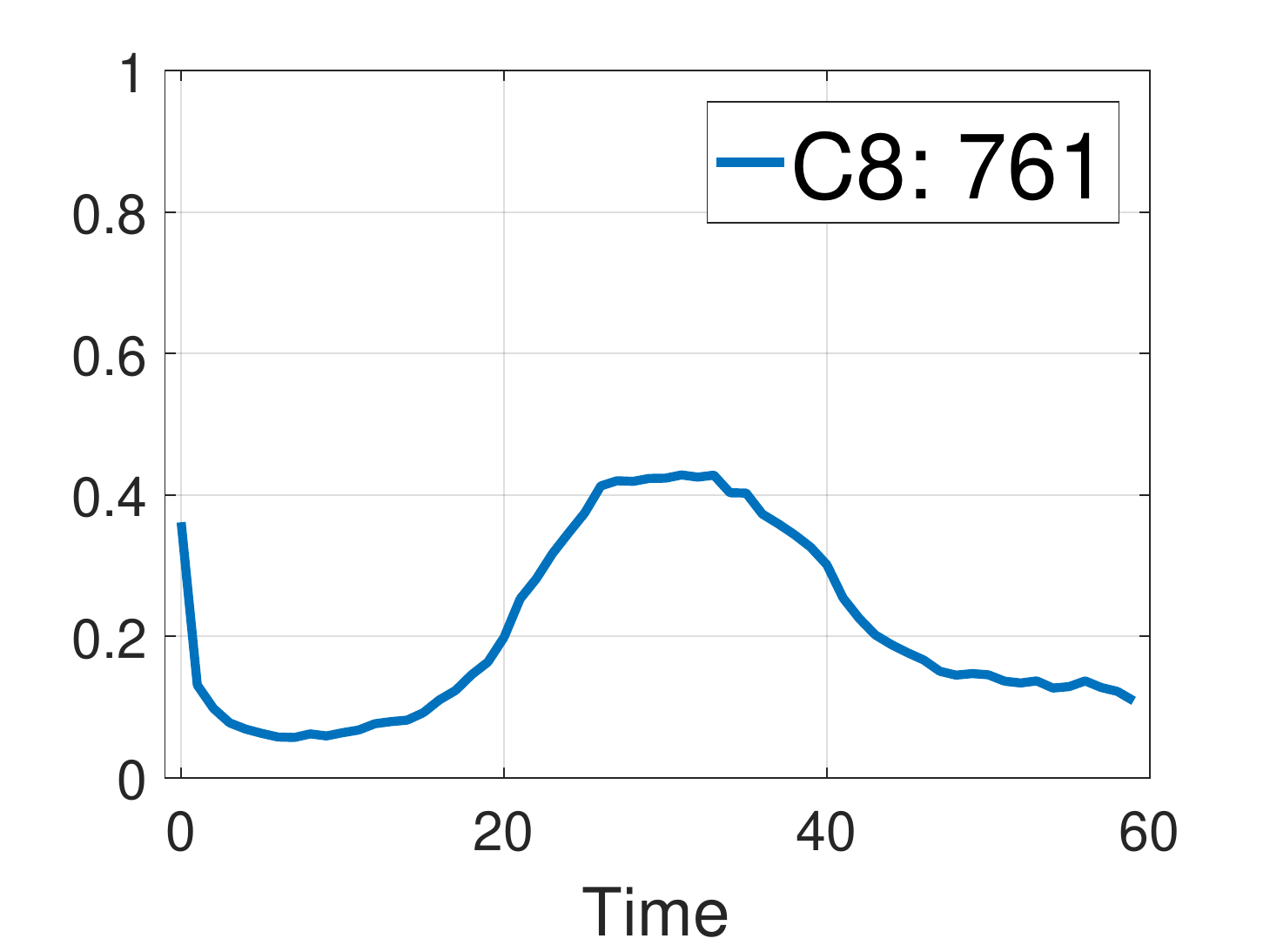}
\label{fig:Intro1}
}
\subfigure[C9: Persist]{
\includegraphics[width=0.147\textwidth, trim = 0 0 0 0, clip]{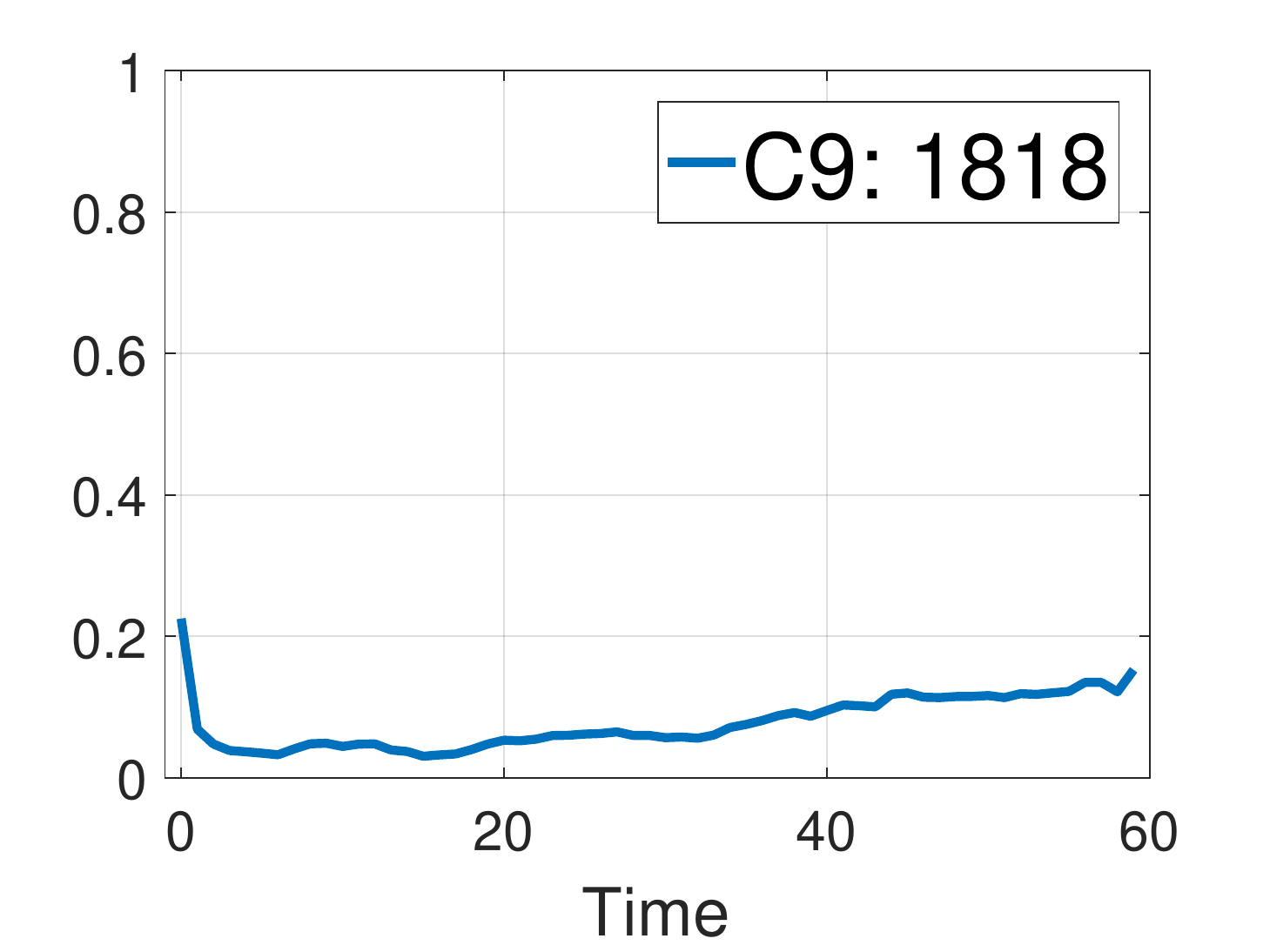}
\label{fig:Intro1}
}
\caption{ {The cluster centroids of empirical cascade dynamics. The number in legend is the number of instances in each cluster.}
\vspace{-0.1in}
\label{fig:temporalCluster}}
\end{figure}

\mytag{Clustering intuitions.} 
To facilitate the analysis of the interplay between cascade dynamics and their  structures, we propose three high level clustering intuitions.
 \begin{compactitem}
\item{Local shape:}  similar local temporal patterns of cascades may correspond to similar local structures. For example, spike patterns (as shown in Fig.~\ref{fig:temporalCluster}a) possibly correspond to star-like structures, while persistent growth patterns (as shown in Fig.~\ref{fig:temporalCluster}i) possibly correspond to deep-tree-like structures.
\item{Position:}  the same shapes at the similar temporal positions over the time dimension may correspond to sub-structures share similar positions in their embedded cascades. For example,  spikes at initial time (Fig.~\ref{fig:temporalCluster}a) possibly correspond to star-like structures centering around the original poster, while latter spikes  (Fig.~\ref{fig:temporalCluster}e) possibly corresponding to star-like structures centering around retweeters at deeper positions.
\item{Global shape:} the dynamics should be captured as a whole rather than focusing only on a local part. For example, the dynamics of cascades are a combinations of rises, falls and their positions. A local spike can only correspond to a star-like sub-cascade as a specific part of the whole cascade.
\end{compactitem}
In addition, the clustering algorithm should be scalable to our large dataset.

\mytag{Clustering methods and results.} Based on the above intuitions, we examine three state-of-the-art time-series clustering methods, i.e., K-SC \cite{yang2011patterns}, K-Shape \cite{paparrizos2015k}, K-Means \cite{kaufman2009finding}. We choose the cascades with mass $\ge 100$ to get relatively smooth dynamic curves. K-SC and K-Shape can capture the shape of dynamics because they are based on the shape similarities considering scaling invariance. However, their linear-shifting-invariant properties contradict our position intuition. In addition, K-SC is not scalable to our data scale. We apply K-Shape to our whole dataset and K-SC to a randomly sampled dataset which consists of $10,000$ temporal dynamics. By checking the clustering centers and corresponding dynamic instances, we find K-SC gives reasonable local shape but ignores the position and thus the global shape. The shape extraction method of K-Shape can not generate reasonable dynamic centers. Finally, in order to capture the above intuitions, we normalize the time of $c(t)$ by cascade lifetime to capture the relative position, scale the $c(t)$ by its maximum value to capture the shape, and choose K-Means to capture the global similarity. 
Figure~\ref{fig:temporalCluster} plots the centroids of nine dynamic clusters we discovered, and we name them by their  shapes, i.e., C1 spike, C2 fat-spike, C3 fatter-spike, C4 small-rebound, C5 big-rebound, C6 late-rebound, C7 early-persist, C8 late-persist, and C9 persist  for Figs.~\ref{fig:temporalCluster}a-i respectively. 

To find the best cluster numbers or the best clusters are still open problems because we don't know what the correctness is. 
For example, it is tricky to decide whether we should combine cascades in spike (Fig.~\ref{fig:temporalCluster}a), fat-spike (Fig.~\ref{fig:temporalCluster}b) and fatter-spike (Fig.~\ref{fig:temporalCluster}c) clusters into one cluster or to divide  them into more fine-grained clusters.
However, we find relatively robust correlations between dynamics and their structures in a statistical sense.  

\subsection{Implications}

\begin{figure}[!htb]
\centering
\subfigure[]{
\includegraphics[width=0.227\textwidth, trim = 0 0 0 0 , clip]{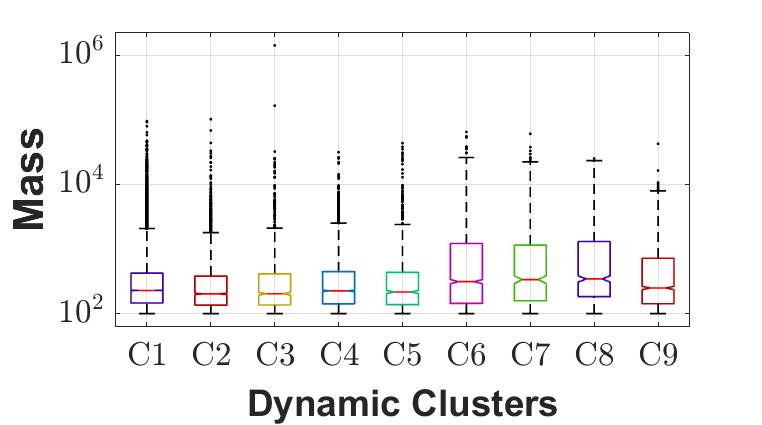}
\label{fig:venn}
}
\subfigure[]{
\includegraphics[width=0.227\textwidth, trim = 0 0 0 0, clip]{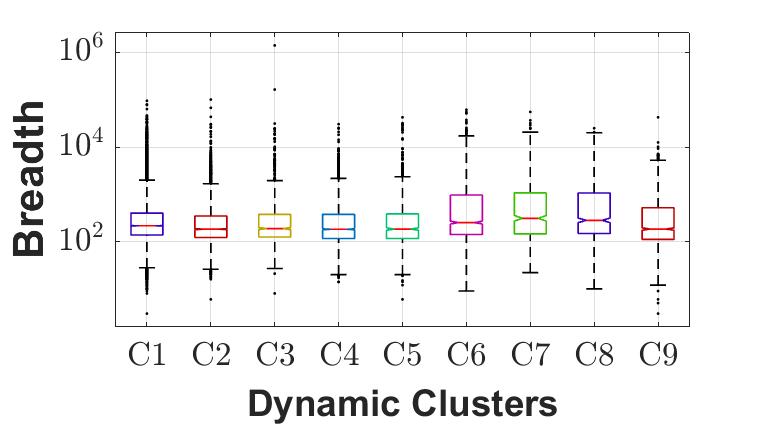}
\label{fig:Intro1}
}
\subfigure[]{
\includegraphics[width=0.227\textwidth, trim = 0 0 0 0, clip]{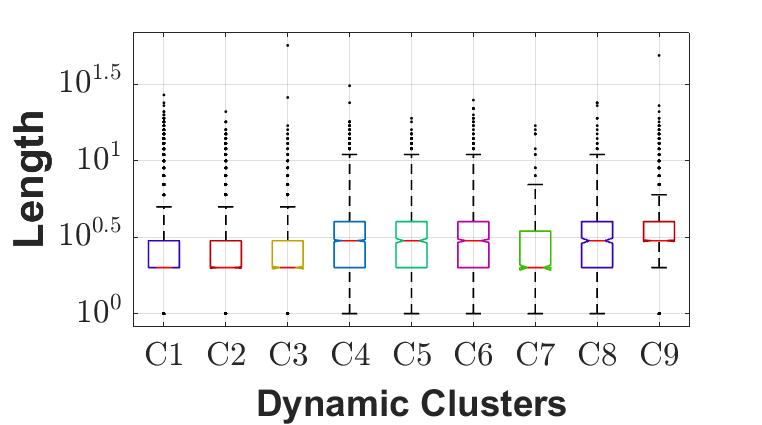}
\label{fig:Intro2}
}
\subfigure[]{
\includegraphics[width=0.227\textwidth, trim = 0 0 0 0, clip]{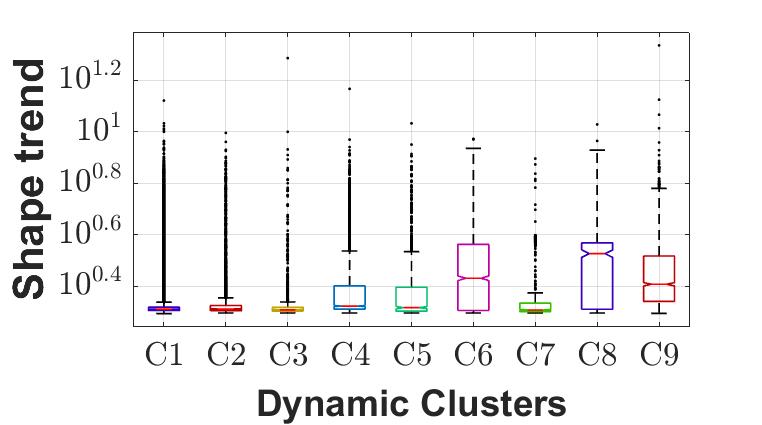}
\label{fig:Intro2}
}
\subfigure[]{
\includegraphics[width=0.227\textwidth, trim = 0 0 0 0, clip]{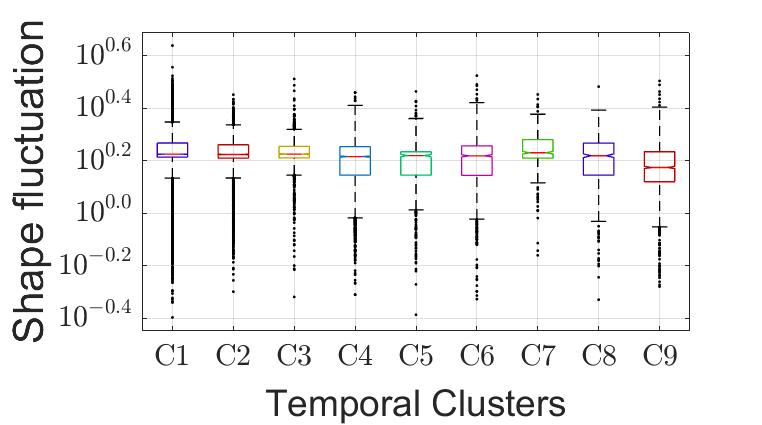}
\label{fig:Intro2}
}
\subfigure[]{
\includegraphics[width=0.227\textwidth, trim = 0 0 0 0, clip]{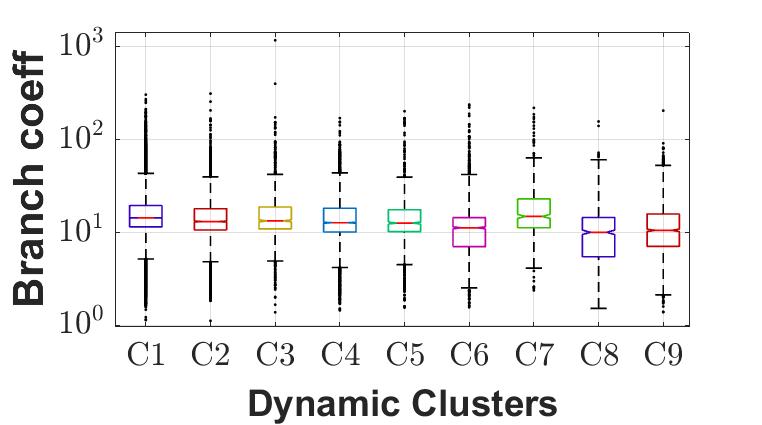}
\label{fig:Intro1}
}
\subfigure[]{
\includegraphics[width=0.227\textwidth, trim = 0 0 0 0, clip]{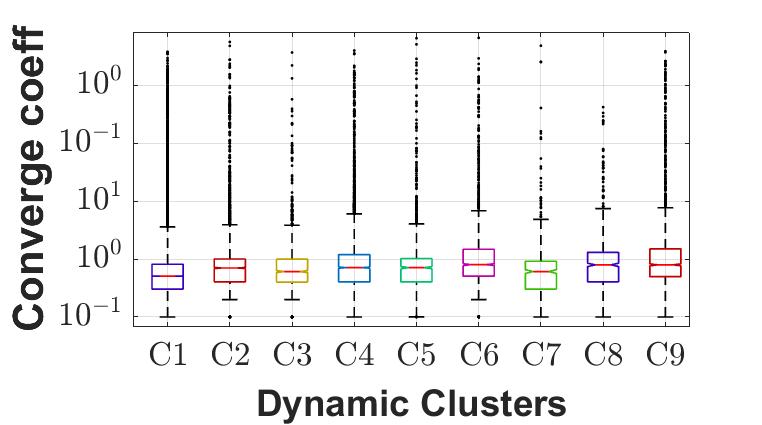}
\label{fig:Intro2}
}
\subfigure[]{
\includegraphics[width=0.227\textwidth, trim = 0 0 0 0, clip]{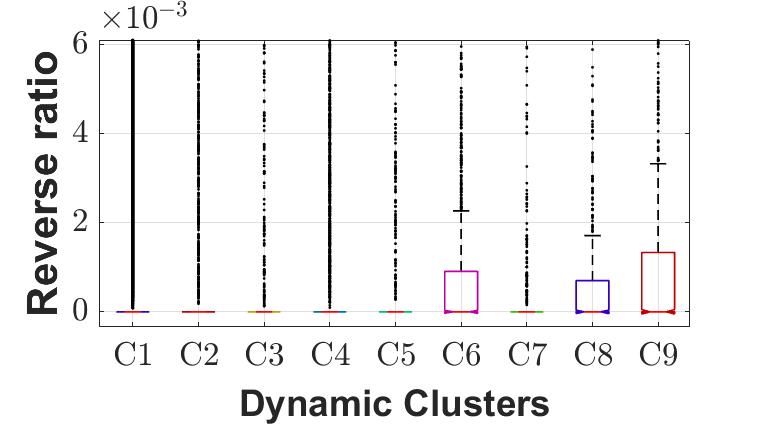}
\label{fig:Intro1}
}
\subfigure[]{
\includegraphics[width=0.227\textwidth, trim = 0 0 0 0, clip]{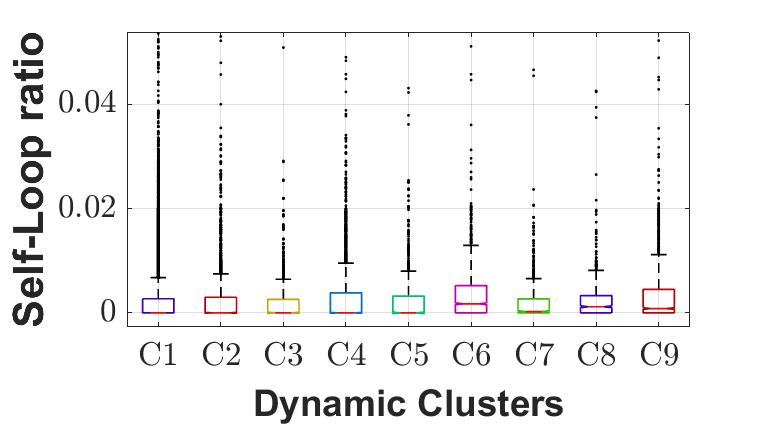}
\label{fig:Intro1}
}
\subfigure[]{
\includegraphics[width=0.227\textwidth, trim = 0 0 0 0, clip]{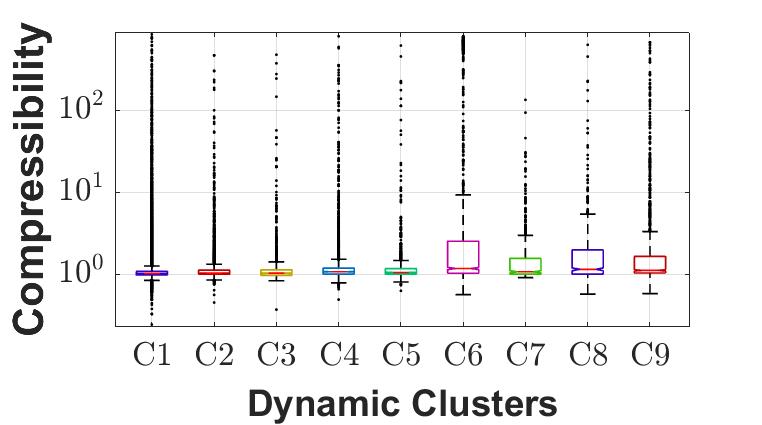}
\label{fig:Intro1}
}
\vspace{-0.2in}
\caption{Structure metric distributions of dynamic clusters represented by boxplots, among which only reciprocity and self-loop ratio are reported in linear scale and all the other metrics are reported by the exponents after being logarithmically transformed, i.e., we compare their differences in their order of magnitude. 
\label{fig:structureOfTemporalCluster}}
\vspace{-0.2in}
\end{figure}

\mytag{Structural distinguishability between dynamic clusters.} 
We first examine whether the structures can distinguish  dynamic clusters. Here, we test the null hypothesis that cascades in different dynamic clusters come from the same structure metric distribution by using Kruskal-Wallis test \cite{kruskal1952use}. We find that null hypothesis for each structure metric as shown in Fig.~\ref{fig:structureOfTemporalCluster} is rejected at the significance level $p<10^{-10}$, indicating that different dynamic clusters do not share same structures, implying that structures can distinguish the dynamics.

\begin{figure}[!ht]
\begin{center}
\centerline{\includegraphics[width=.35\textwidth]{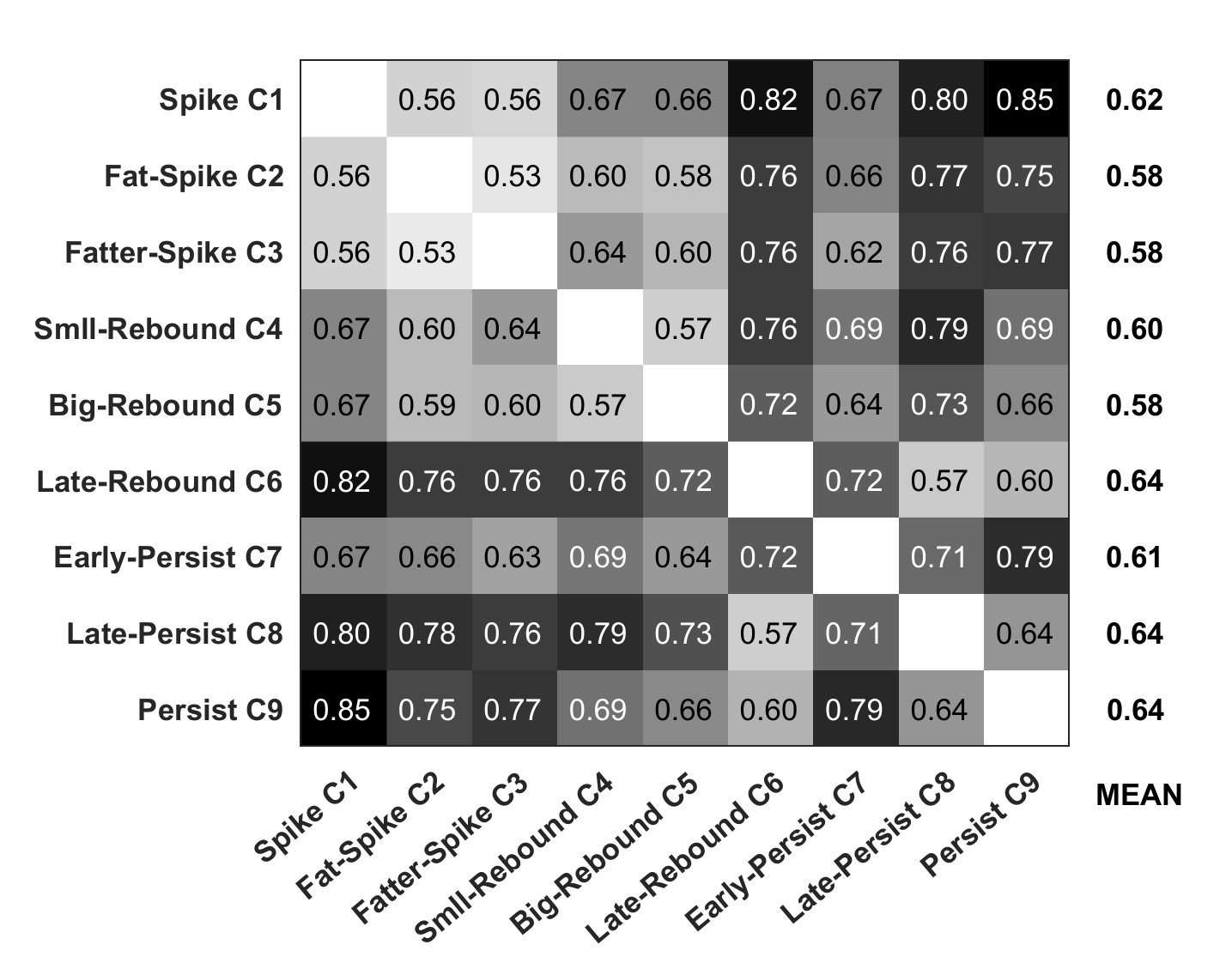}}
\caption{{Structural distinguishability between dynamic clusters.} 
}\label{fig:sdDynamics}
\end{center}
\vspace{-0.4in}
\end{figure}  

However, to what extent can the structures distinguish  different dynamic clusters?  In order to quantify the distinguishability of cascade structures in dynamics, e.g., cluster C1 and cluster C2, we use the best classification accuracy of a linear classifier applied in a balanced number of cascades in C1 and C2 to quantify this structural distinguishability between dynamic clusters. Thus, the random guessing results are served as a baseline with accuracy $0.5$. Figure~\ref{fig:sdDynamics} plots the structural distinguishability between each pair of dynamic clusters. We find distinguishability values between  C1 vs. C2, C1 vs. C3 and C2 vs. C3 are $0.56$, $0.56$ and $0.53$ respectively, indicating that although C1 ,C2 and C3 show distinguishable decay in dynamic dimension as shown in Figs.~\ref{fig:temporalCluster}a-c, they share quite similar (indistinguishable ) structures. In contrast, C6, C8 and C9 in Fig.~\ref{fig:sdDynamics} have large distinguishability values with other clusters, indicating that structures can distinguish their dynamics.  In all, we find structures can distinguish different dynamics with different distinguishability values.

After examining the structure metrics as a whole black box, we next analyze interpretable implications of structures for dynamics by answering following four questions.

\mytag{Which dynamic patterns are bigger in mass, longer in length, and wider in breadth?}
We find that although the cascade with largest mass value belongs to C3 fatter-spike, the majority of cascades in C3 are not large in mass, as shown in Fig.~\ref{fig:structureOfTemporalCluster}a. In contrast, cascades in C6-9 are usually large in mass. For instance, C8 late-persist has the largest median value of mass because their dynamic patterns exhibit a lasting high growth rate at the second half of their lifetime. Similar results are applied to the breadth metrics in Fig.~\ref{fig:structureOfTemporalCluster}b. As for the length in Fig.~\ref{fig:structureOfTemporalCluster}c, we find cascades in C4-6 (small-rebound, big-rebound, late-rebound) and C8-9 (late-persist, and persist) show longer length values. Indeed, we find all these dynamics patterns show a second growth in their lifetime just as their names suggest. 

\mytag{Which dynamic patterns are viral in structures?}
Previous works used the Wiener index to measure the structural virality of cascades \cite{goel2015structural}. We now analyze the the correlations between viral cascades and their corresponding dynamics. Figures~\ref{fig:structureOfTemporalCluster}d\&e plot the distributions of silhouette trend values (measured by the Wiener index) and silhouette fluctuation values for each dynamics clusters. We find spike patterns in C1-C3 show small trend values in  Fig.~\ref{fig:structureOfTemporalCluster}d, implying that star-like graphs are accompanied by spike dominated dynamics patterns. 
In contrast, C6 late-rebound, C8 late-persist and C9 persist patterns exhibit large trend values, implying that cascades with deep structures also exhibit long lasting growth patterns in their lifetime, especially at the second half of their lifetime. At the same time, compared with C9, cascades in C6 and C8 show relatively larger fluctuation values  as shown in  Fig.~\ref{fig:structureOfTemporalCluster}e, indicating that cascades in C6 and C8 tend to have larger breadth values at some depth while cascades in C9 tend to be about the same breadth values at each depth. These structures of C6, C8 and C9 also correspond to their temporal dynamics. For example, C6 and C8 exhibit relatively larger growth rate while C9 tends to grow gradually without obvious spikes.

\mytag{Which dynamic patterns show polarized spreading directions?}
Figures~\ref{fig:structureOfTemporalCluster}f-i plot the distributions of direction metrics for each dynamics clusters. We find cascades in C7 early-persist have the largest median branch coefficient as shown in Fig.~\ref{fig:structureOfTemporalCluster}f, indicating the existence of a couple of hubs where informations mainly spread out from them. The early-persist pattern of C7 implies that the lasting growth at the early time are due to the retweeters directly infected by these one or two hubs which leads to large branch value. In contrast, C6 late-rebound,  C8 later-persist and C9 persist show smallest median branch value, and have a obvious latter growth (rebound growth  in Fig.~\ref{fig:temporalCluster}f and persist growth in Fig.~\ref{fig:temporalCluster}h), implying that the latter growth is due to the spreads from a lot of already infected users rather than few hubs and thus have small branch coefficient.

In addition, the latter growth patterns in C6, C8 and C9 also have large converge values (Fig. ~\ref{fig:structureOfTemporalCluster}g), reciprocity (Fig. ~\ref{fig:structureOfTemporalCluster}h), and self-loop ratios (Fig. ~\ref{fig:structureOfTemporalCluster}i), indicating that information flows in these patterns tend to flow into few users, flow reversely and just stay with the spreaders,
implying  that the intensive interactions and manipulations of information flows lead to these latter growth patterns and thus viral structures.

\mytag{Which dynamic patterns show borrowed prosperity?} We find large average activity values for C6-C9 in Fig.~\ref{fig:structureOfTemporalCluster}j, especially for cascades in C6 late-rebound and C8 late-persist, implying the existence of a large amount of repeated retweets and promotions. The rationalities behind the C6 and C8 may lay in the fact that an initial unpopular post are popular later due to the frequent retweets and promotions by promoters. 

The repeated promotions can also account for dynamics patterns C7 early-persist, implying the repeated promotions leads to the persistent early growth, where the growth is dominated by new infected users directly retweeting from few hubs as shown in Fig.~\ref{fig:structureOfTemporalCluster}f with large branch value and small trend value in  Fig.~\ref{fig:structureOfTemporalCluster}d. In contrast, temporal pattern C9 (persist in Fig.~\ref{fig:temporalCluster}i) also needs repeated promotions to facilitate growth, where the growth is dominated by new infected users who retweets from a crowd of already existing users (small branch value and large trend value) rather than a couple of hubs.

\section{Implications for semantics}
It remains a largely unknown question that how the semantics of a cascade interplays with its structure.
Each microblog (or weibo) has a specific semantic expressed by a set of multi-modal informations including the text, embedded hashtags, attached pictures or videos, the poster and his/her profile tags and self-descriptions. We characterize the semantics of a microblog by topic modeling based on above informations, and  then investigate the correlations between cascade semantics and its corresponding structures.


\subsection{Characterizing cascade semantics}
In order to get the topic label with high precision, we adopt filtering methods discussed in Refs.~\cite{romero2011differences,yang2014large}. 
We first build the initial taxonomy of topics to describe the semantics of microblogs. Starting for the existing taxonomy of topics shown in Refs. \cite{romero2011differences,yang2014large}, we update the taxonomy with new-found topics which do not fit in our already-existed taxonomy during the following procedures. We iteratively use hashtag priors and user priors to filter the microblogs and followed by the manual inspection to update topic priors. Specifically, 
we first extract and sort hashtags in our dataset and manually label the unambiguous hashtags (e.g., \#animation\#, \#movie\#) according to the taxonomy as hashtag priors for filtering.  For the user priors, we first sample a set of seed users (the number of verified users vs. ordinary users $ \approx 1:1$) with user names, profile tags, self-descriptions and all the hashtags in his/her posts, and then manually assign one unique topic to these users by examining above informations. For example, verified users like QQMusic or QQFilm consistently published the topics as their names suggest. Users with more than one topic are omitted. All the microblogs posted by these seed users are labeled with their authors' topic. Manually inspections are applied to validate whether the contents of the filtered microblogs are consistent with the topic priors. 
Due to the prevalance of non-topical microblogs, we cover a subset of cascades in our dataset which have topic labels with relatively high precision.
In all, we find $14$ major topics as shown in Table~\ref{tab:topic}. 

\begin{table}[!htbp]
\vspace{-0.2in}
\scriptsize
\begin{center}
\caption{Discovered Topics\label{tab:topic}}
\label{tab:Symbol}
\begin{tabular}{c|c|c|c|c|c}
\hline  
\hline  
\textbf{Topic} &Livings& Funny & Travel & Celebrity & Animation\\ 
\textbf{No.} & 48,230  & 20,710 & 2,046 & 1,307 &  840\\ 
\hline
\textbf{Topic} &Visual Arts& Technology & Politics & Business & Movies\\ 
\textbf{No.} & 794  & 731 & 540 & 401 & 295 \\ 
\hline 
\textbf{Topic} &Music& Games & Sports & Raffle& \\ 
\textbf{No.} & 289  & 212 & 195 & 123 &  \\ 
\hline
\hline
\end{tabular} 
\end{center}
\vspace{-0.2in}
\end{table}


\subsection{Implications}

\mytag{Structural distinguishability between topics.} 
Following the same method discussed in Section 5.2, we find that structures can distinguish topics by testing and rejecting (at the significance level $p<10^{-10}$) the null hypothesis that cascades with different topics share same structure metric distributions. 

\begin{figure}[!ht]
\vspace{-0.1in}
\begin{center}
\centerline{\includegraphics[width=.35\textwidth]{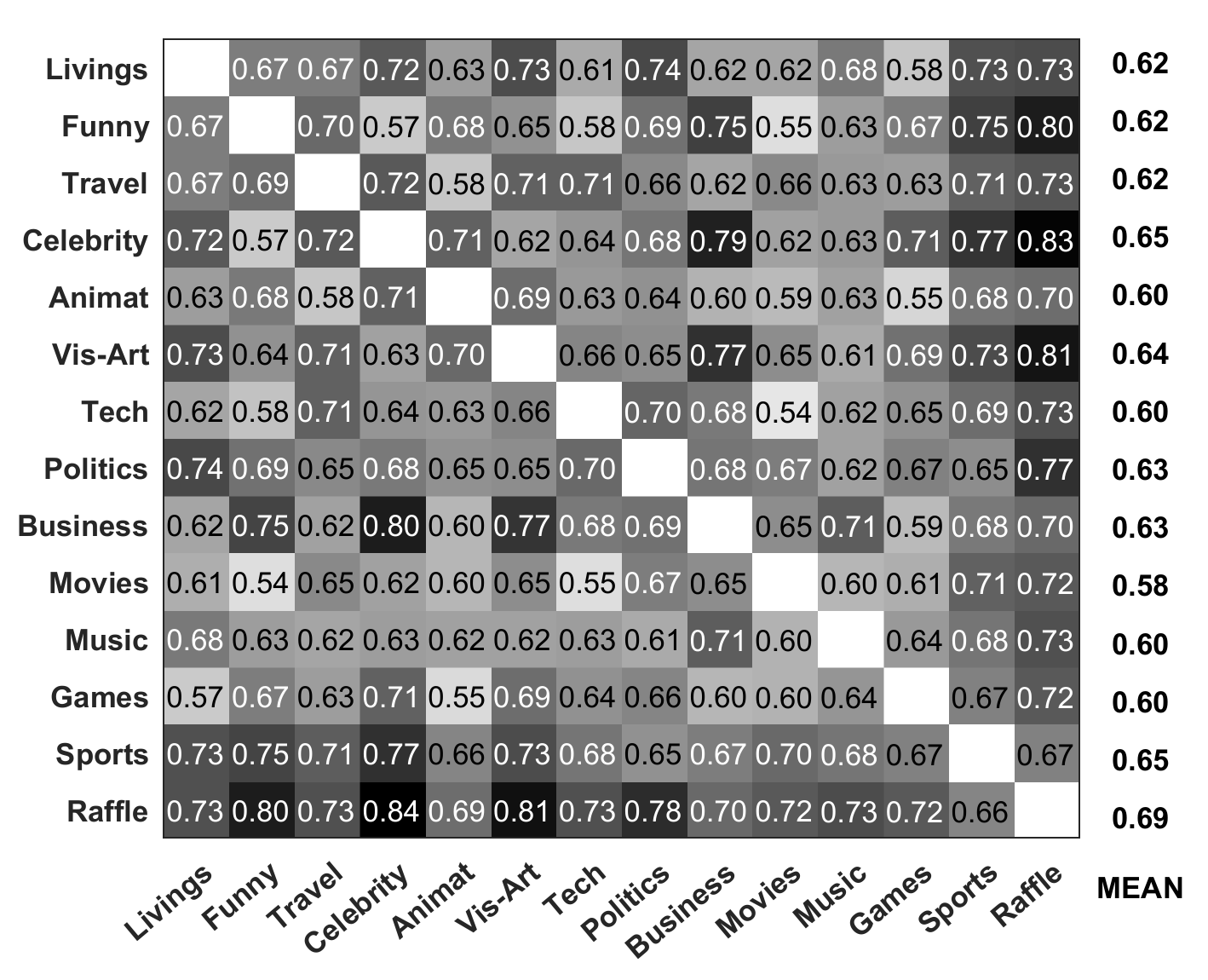}}
\vspace{-0.1in}
\caption{{Structural distinguishability between topics.} 
}
\label{fig:sdTopics}
\end{center}
\vspace{-0.3in}
\end{figure}  

We then try to answer to what extent structures can distinguish topics. Figure~\ref{fig:sdTopics} plots the pair-wise distinguishability of structures in each pair of topics. First, we find the average structural distinguishability  between topics (Fig.~\ref{fig:sdTopics}) and between dynamics (Fig.~\ref{fig:sdDynamics}) are $0.62$ and $0.61$ respectively, implying the distinguishability of structures for topics is a little better than (or at least as good as) structures for dynamics in average. Second, different topics have different distinguishability values. For instance, the raffle topic is the most distinguishable topic (mean distinguishability value $0.69$) while movie is the most indistinguishable topic (mean distinguishability value $0.58$). Third, different topic pairs also have different discriminability values. For instance, raffle and celebrity are the most distinguishable two topics (distinguishability value $0.84$). In contrast, technology and movies are almost indistinguishable (distinguishability value $0.54$).

\begin{figure}[htb]
\centering
\subfigure[]{
\includegraphics[width=0.227\textwidth, trim = 0 0 0 0 , clip]{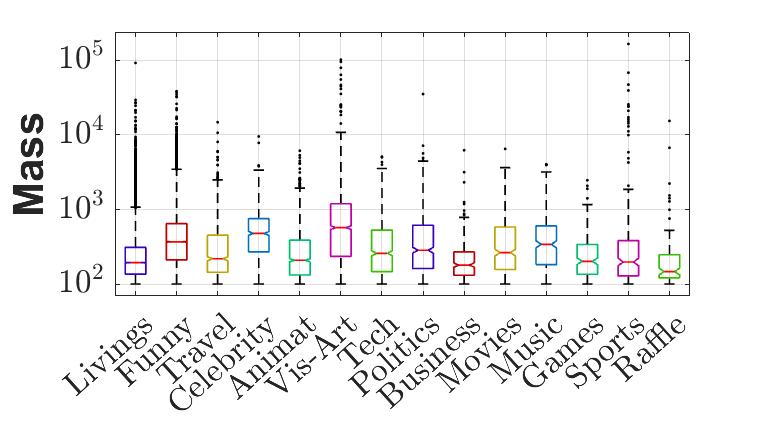}
\label{fig:venn}
}
\subfigure[]{
\includegraphics[width=0.227\textwidth, trim = 0 0 0 0, clip]{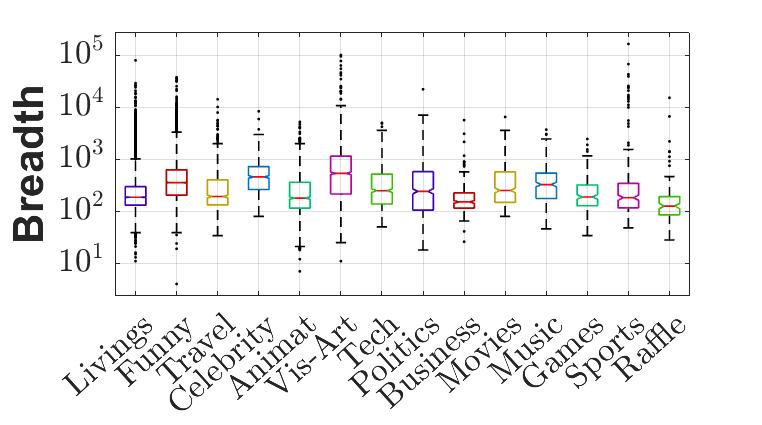}
\label{fig:Intro1}
}
\subfigure[]{
\includegraphics[width=0.227\textwidth, trim = 0 0 0 0, clip]{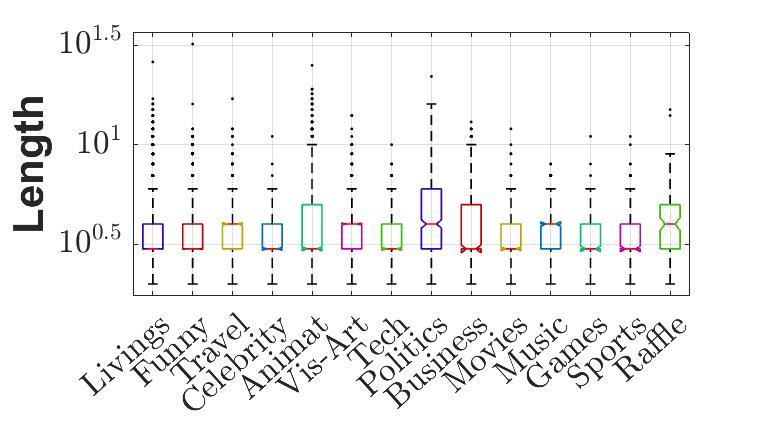}
\label{fig:Intro2}
}
\subfigure[]{
\includegraphics[width=0.227\textwidth, trim = 0 0 0 0, clip]{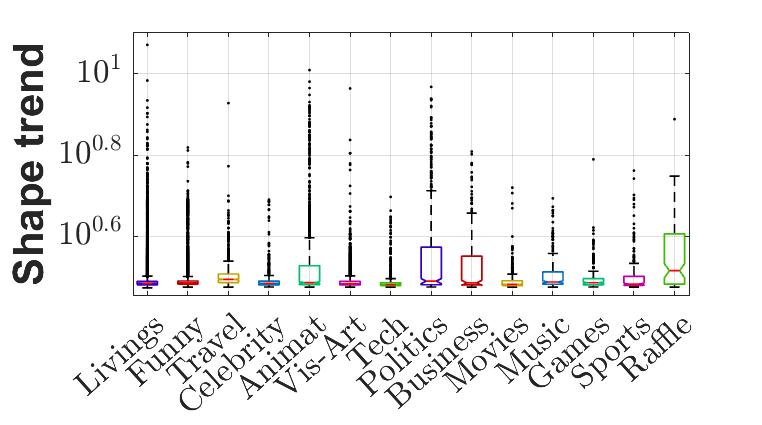}
\label{fig:Intro2}
}
\subfigure[]{
\includegraphics[width=0.227\textwidth, trim = 0 0 0 0, clip]{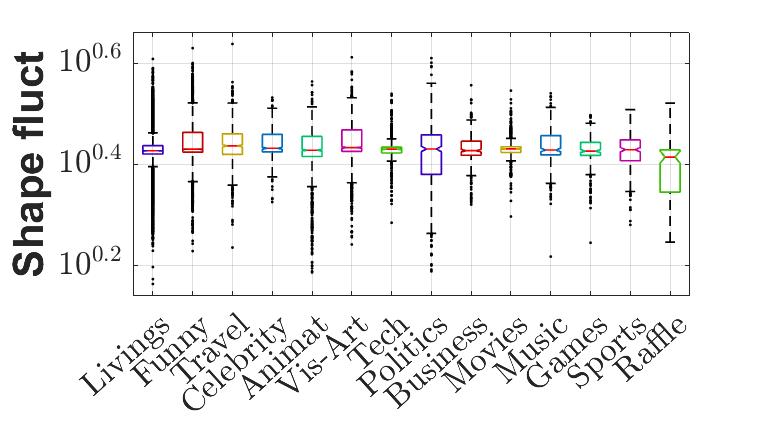}
\label{fig:Intro2}
}
\subfigure[]{
\includegraphics[width=0.227\textwidth, trim = 0 0 0 0, clip]{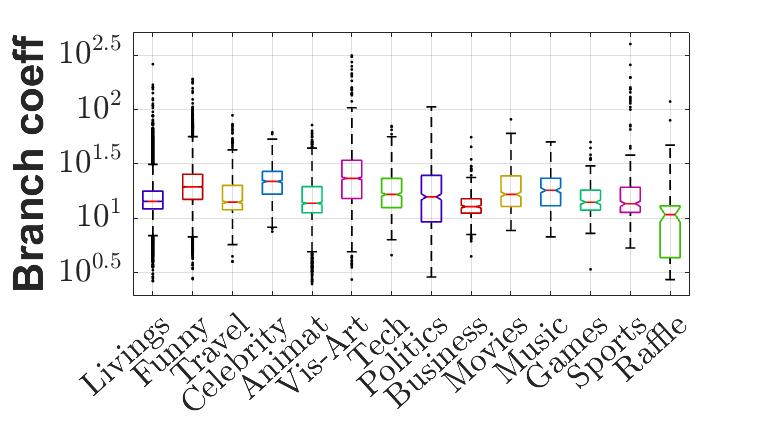}
\label{fig:Intro1}
}
\subfigure[]{
\includegraphics[width=0.227\textwidth, trim = 0 0 0 0, clip]{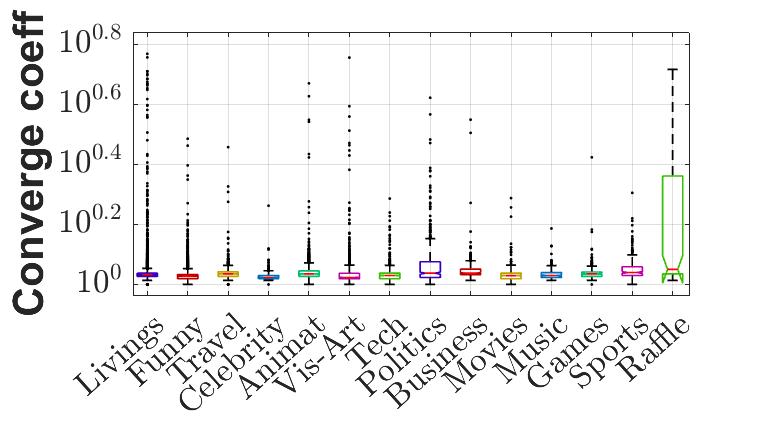}
\label{fig:Intro2}
}
\subfigure[]{
\includegraphics[width=0.227\textwidth, trim = 0 0 0 0, clip]{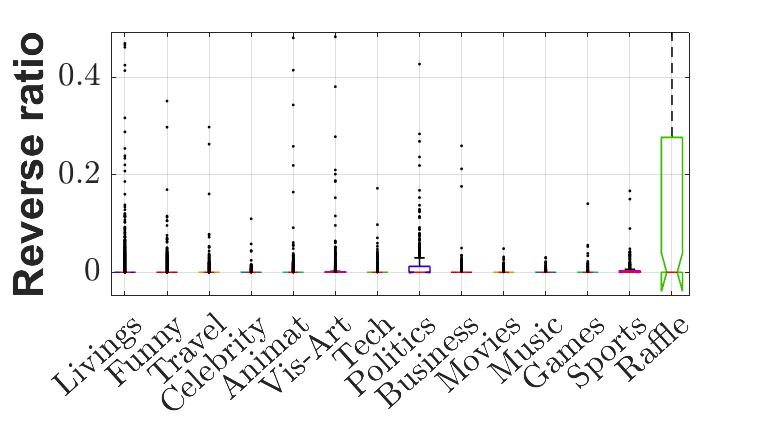}
\label{fig:Intro1}
}
\subfigure[]{
\includegraphics[width=0.227\textwidth, trim = 0 0 0 0, clip]{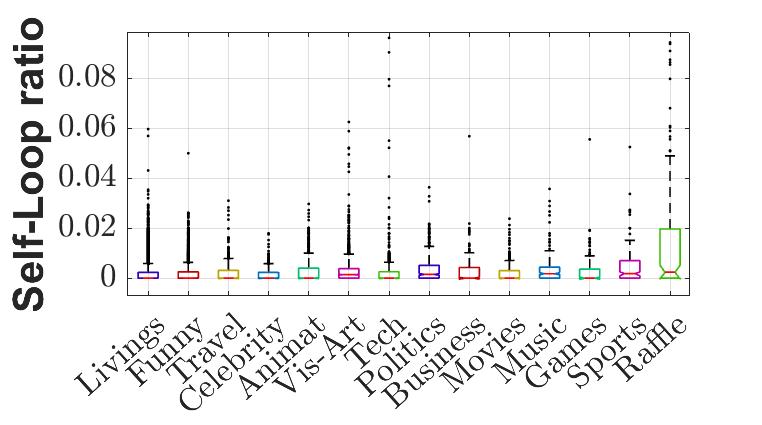}
\label{fig:Intro1}
}
\subfigure[]{
\includegraphics[width=0.227\textwidth, trim = 0 0 0 0, clip]{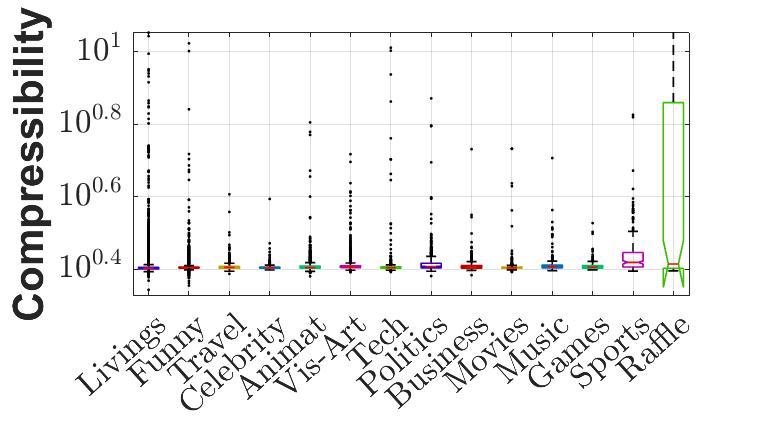}
\label{fig:Intro1}
}
\vspace{-0.2in}
\caption{ {Structure metric distributions of topics.}
\label{fig:structureOfTopic}}
\vspace{-0.2in}
\end{figure} 

Figure~\ref{fig:structureOfTopic} plots the distributions (represented by box-plot) of structure metrics for each topic.  Again, we analyze the structural differences between topics by answering following four questions.

\mytag{Which topics are large in size?}
We find visual art topic tends to have large mass (Fig.~\ref{fig:structureOfTopic}a) and breadth (Fig.~\ref{fig:structureOfTopic}b), implying that the microblogs with beautiful pictures can attract a lot of audiences. As for the length (Fig.~\ref{fig:structureOfTopic}c), politics topic tends to spread further.

\mytag{Which topics are viral in structures?}
We find politics, business, and raffle topics have large silhouette values (Fig.~\ref{fig:structureOfTopic}d) and moderate fluctuation values (Fig.~\ref{fig:structureOfTopic}e), implying these topics tend to spread deeper and the numbers of infected users at each depth are about the same, i.e., narrow and deep patterns. In contrast, topics like livings, celebrity tend to follow wide and shallow patterns.

\mytag{Which topics show polarized spreading directions?}
Raffle topic exhibits small branch coefficient (Fig.~\ref{fig:structureOfTopic}f), but with large converge coefficient (Fig.~\ref{fig:structureOfTopic}g), reverse ratio (Fig.~\ref{fig:structureOfTopic}h), self-loop ratio (Fig.~\ref{fig:structureOfTopic}i), due to the fact that involved users in raffle topic want to raise their probability of being awarded by manipulating information spreadings. Politics topic also exhibit relatively large converge deviation value, reciprocity, self-loop ratio possibly due to the fact that involved users are discussing controversial political issues.

\mytag{Which topics show borrowed prosperity?} No doubt, it is raffle topic (Fig.~\ref{fig:structureOfTopic}j). However,sports topic also exhibit relatively large average activity values may due to the fact that involved users actively spread information about their concerned sport events or their favorite sports stars.





%
 
\section{Conclusions}
\label{sec:concl}
In this paper, we analyze and quantify the structural patterns of information cascades and the interplay  between structures, dynamics and semantics. We collect the full scale information cascades generated during one week in Tencent Weibo to support our detailed empirical studies. 
In order to quantify the complex structure pattern of information cascades, we propose a ten-dimensional structural metric to reflect the size, silhouette, direction and activity aspects of cascades. 
We find: (a) the bimodal distribution governs the mass, length, and breadth, wiener index and the number of reciprocal edges;
(b) information flows in empirical cascades have four directions, namely branching-out, converging-in, reciprocal, and self-loop;
(c) repeated retweets and self-promotions are prevalent. The average activity and the number of self-loop edges follow power law.
We further study the high-order structural patterns of cascades by answering: (a) Are cascade wide and shallow? Or narrow and deep? (b) To what extent do cascades follow the star-like pattern or chain-like pattern? (c) To what extent do the four directions of information flow coexist in a cascade? (d) What are the structure patterns of so-called popular cascades?


Based on the cascade structure analysis, we further investigate their correlations with the dynamics and semantics. We first evaluate to what extent  structural features of a information cascade can explain its corresponding dynamics and topics. The results show that the structures of information cascades have notable correlations with its dynamics and semantics. We then conduct more insightful case studies by answering following questions: (a) Which dynamic/semantic patterns are bigger in mass, longer in length and wider in breadth? (b) Which dynamic/semantic patterns are viral in structure? (c) Which dynamic/semantic patterns show polarized spreading directions? (d) Which dynamic/semantic patterns show borrowed prosperity?

\mytag{Limitations and further works.} It's worthwhile to validate our findings in other datasets like twitter. Though we propose four aspects of cascade structure, more elaborate metrics accounting for these aspects should be examined. More elaborate methods on modeling the dynamics clusters and topics need further investigation.

\section*{Acknowledgments}
\small{
This work was supported by National Program on Key Basic Research Project, No. 2015CB352300; National Natural Science Foundation of China, No. 61370022, No. 61531006, No. 61472444 and No. 61210008. Thanks for the research fund of Tsinghua-Tencent Joint Laboratory for Internet Innovation Technology.
This material is based upon work
   supported by the National Science Foundation
   under Grant No.
   CNS-1314632 
   IIS-1408924 
    Any opinions, findings, and conclusions or recommendations expressed in this
    material are those of the author(s) and do not necessarily reflect the views
    of the National Science Foundation, or other funding parties.
}
 
\bibliographystyle{abbrv}
\small
\bibliography{sigproc}  

\begin{thebibliography}{10}

\bibitem{adar2004implicit}
E.~Adar, L.~Zhang, L.~A. Adamic, and R.~M. Lukose.
\newblock Implicit structure and the dynamics of blogspace.
\newblock In {\em Workshop on the weblogging ecosystem}, volume~13, pages
  16989--16995, 2004.

\bibitem{anderson2015global}
A.~Anderson, D.~Huttenlocher, J.~Kleinberg, J.~Leskovec, and M.~Tiwari.
\newblock Global diffusion via cascading invitations: Structure, growth, and
  homophily.
\newblock In {\em Proceedings of the 24th International Conference on World
  Wide Web}, pages 66--76. ACM, 2015.

\bibitem{blei2012probabilistic}
D.~M. Blei.
\newblock Probabilistic topic models.
\newblock {\em Communications of the ACM}, 55(4):77--84, 2012.

\bibitem{cheng2014can}
J.~Cheng, L.~Adamic, P.~A. Dow, J.~M. Kleinberg, and J.~Leskovec.
\newblock Can cascades be predicted?
\newblock In {\em Proceedings of the 23rd international conference on World
  wide web}, pages 925--936. ACM, 2014.

\bibitem{cheng2016cascades}
J.~Cheng, L.~A. Adamic, J.~M. Kleinberg, and J.~Leskovec.
\newblock Do cascades recur?
\newblock In {\em Proceedings of the 25th International Conference on World
  Wide Web}, pages 671--681. International World Wide Web Conferences Steering
  Committee, 2016.

\bibitem{dow2013anatomy}
P.~A. Dow, L.~A. Adamic, and A.~Friggeri.
\newblock The anatomy of large facebook cascades.
\newblock In {\em ICWSM}, 2013.

\bibitem{goel2015structural}
S.~Goel, A.~Anderson, J.~Hofman, and D.~J. Watts.
\newblock The structural virality of online diffusion.
\newblock {\em Management Science}, 62(1):180--196, 2015.

\bibitem{goel2012structure}
S.~Goel, D.~J. Watts, and D.~G. Goldstein.
\newblock The structure of online diffusion networks.
\newblock In {\em Proceedings of the 13th ACM conference on electronic
  commerce}, pages 623--638. ACM, 2012.

\bibitem{hong2010empirical}
L.~Hong and B.~D. Davison.
\newblock Empirical study of topic modeling in twitter.
\newblock In {\em Proceedings of the first workshop on social media analytics},
  pages 80--88. ACM, 2010.

\bibitem{kaufman2009finding}
L.~Kaufman and P.~J. Rousseeuw.
\newblock {\em Finding groups in data: an introduction to cluster analysis},
  volume 344.
\newblock John Wiley \& Sons, 2009.

\bibitem{kossinets2008structure}
G.~Kossinets, J.~Kleinberg, and D.~Watts.
\newblock The structure of information pathways in a social communication
  network.
\newblock In {\em Proceedings of the 14th ACM SIGKDD}. ACM, 2008.

\bibitem{kruskal1952use}
W.~H. Kruskal and W.~A. Wallis.
\newblock Use of ranks in one-criterion variance analysis.
\newblock {\em Journal of the American statistical Association},
  47(260):583--621, 1952.

\bibitem{leskovec2007dynamics}
J.~Leskovec, L.~A. Adamic, and B.~A. Huberman.
\newblock The dynamics of viral marketing.
\newblock {\em ACM Transactions on the Web (TWEB)}, 1(1):5, 2007.

\bibitem{leskovec2007patterns}
J.~Leskovec, M.~McGlohon, C.~Faloutsos, N.~S. Glance, and M.~Hurst.
\newblock Patterns of cascading behavior in large blog graphs.
\newblock In {\em SDM}, volume~7, pages 551--556. SIAM, 2007.

\bibitem{liben2008tracing}
D.~Liben-Nowell and J.~Kleinberg.
\newblock Tracing information flow on a global scale using internet
  chain-letter data.
\newblock {\em Proceedings of the National Academy of Sciences},
  105(12):4633--4638, 2008.

\bibitem{marquardt1963algorithm}
D.~W. Marquardt.
\newblock An algorithm for least-squares estimation of nonlinear parameters.
\newblock {\em Journal of the society for Industrial and Applied Mathematics},
  11(2):431--441, 1963.

\bibitem{martin2016exploring}
T.~Martin, J.~M. Hofman, A.~Sharma, A.~Anderson, and D.~J. Watts.
\newblock Exploring limits to prediction in complex social systems.
\newblock In {\em Proceedings of the 25th International Conference on World
  Wide Web}, pages 683--694. International World Wide Web Conferences Steering
  Committee, 2016.

\bibitem{matsubara2012rise}
Y.~Matsubara, Y.~Sakurai, B.~A. Prakash, L.~Li, and C.~Faloutsos.
\newblock Rise and fall patterns of information diffusion: model and
  implications.
\newblock In {\em Proceedings of the 18th ACM SIGKDD}. ACM, 2012.

\bibitem{myers2014bursty}
S.~A. Myers and J.~Leskovec.
\newblock The bursty dynamics of the twitter information network.
\newblock In {\em Proceedings of the 23rd international conference on World
  wide web}, pages 913--924. ACM, 2014.

\bibitem{paparrizos2015k}
J.~Paparrizos and L.~Gravano.
\newblock k-shape: Efficient and accurate clustering of time series.
\newblock In {\em Proceedings of the 2015 ACM SIGMOD International Conference
  on Management of Data}, pages 1855--1870. ACM, 2015.

\bibitem{pastor2015epidemic}
R.~Pastor-Satorras, C.~Castellano, P.~Van~Mieghem, and A.~Vespignani.
\newblock Epidemic processes in complex networks.
\newblock {\em Reviews of modern physics}, 87(3):925, 2015.

\bibitem{pei2015exploring}
S.~Pei, L.~Muchnik, S.~Tang, Z.~Zheng, and H.~A. Makse.
\newblock Exploring the complex pattern of information spreading in online blog
  communities.
\newblock {\em PloS one}, 10(5):e0126894, 2015.

\bibitem{rogers2010diffusion}
E.~M. Rogers.
\newblock {\em Diffusion of innovations}.
\newblock Simon and Schuster, 2010.

\bibitem{romero2011differences}
D.~M. Romero, B.~Meeder, and J.~Kleinberg.
\newblock Differences in the mechanics of information diffusion across topics:
  idioms, political hashtags, and complex contagion on twitter.
\newblock In {\em Proceedings of the 20th international conference on World
  wide web}. ACM, 2011.

\bibitem{yang2011patterns}
J.~Yang and J.~Leskovec.
\newblock Patterns of temporal variation in online media.
\newblock In {\em Proceedings of the fourth ACM international conference on Web
  search and data mining}, pages 177--186. ACM, 2011.

\bibitem{yang2014large}
S.-H. Yang, A.~Kolcz, A.~Schlaikjer, and P.~Gupta.
\newblock Large-scale high-precision topic modeling on twitter.
\newblock In {\em Proceedings of the 20th ACM SIGKDD}. ACM, 2014.

\bibitem{ye2010measuring}
S.~Ye and S.~F. Wu.
\newblock Measuring message propagation and social influence on twitter. com.
\newblock In {\em International Conference on Social Informatics}, pages
  216--231. Springer, 2010.

\bibitem{yu2015micro}
L.~Yu, P.~Cui, F.~Wang, C.~Song, and S.~Yang.
\newblock From micro to macro: Uncovering and predicting information cascading
  process with behavioral dynamics.
\newblock In {\em Data mining (ICDM), 2015 IEEE international conference on}.
  IEEE, 2015.

\bibitem{Zang2017}
C.~Zang, P.~Cui, S.~Chaoming, and W.~Zhu.
\newblock Uncover pattern formation of information flow.
\newblock {\em Under review}.

\bibitem{Zang2016}
C.~Zang, P.~Cui, and C.~Faloutsos.
\newblock Beyond sigmoids: The nettide model for social network growth, and its
  applications.
\newblock In {\em Proceedings of the 22Nd ACM SIGKDD}, KDD '16. ACM, 2016.

\bibitem{zhang2016multiscale}
T.~Zhang, P.~Cui, C.~Song, W.~Zhu, and S.~Yang.
\newblock A multiscale survival process for modeling human activity patterns.
\newblock {\em PloS one}, 11(3):e0151473, 2016.

\end{thebibliography}
%
%


\hide{
\appendix
\section{Headings in Appendices}
The rules about hierarchical headings discussed above for
the body of the article are different in the appendices.
In the \textbf{appendix} environment, the command
\textbf{section} is used to
indicate the start of each Appendix, with alphabetic order
designation (i.e. the first is A, the second B, etc.) and
a title (if you include one).  So, if you need
hierarchical structure
\textit{within} an Appendix, start with \textbf{subsection} as the
highest level. Here is an outline of the body of this
document in Appendix-appropriate form:
\subsection{Introduction}
\subsection{The Body of the Paper}
\subsubsection{Type Changes and  Special Characters}
\subsubsection{Math Equations}
\paragraph{Inline (In-text) Equations}
\paragraph{Display Equations}
\subsubsection{Citations}
\subsubsection{Tables}
\subsubsection{Figures}
\subsubsection{Theorem-like Constructs}
\subsubsection*{A Caveat for the \TeX\ Expert}
\subsection{Conclusions}
\subsection{Acknowledgments}
\subsection{Additional Authors}
This section is inserted by \LaTeX; you do not insert it.
You just add the names and information in the
\texttt{{\char'134}additionalauthors} command at the start
of the document.
\subsection{References}
Generated by bibtex from your ~.bib file.  Run latex,
then bibtex, then latex twice (to resolve references)
to create the ~.bbl file.  Insert that ~.bbl file into
the .tex source file and comment out
the command \texttt{{\char'134}thebibliography}.
\section{More Help for the Hardy}
The sig-alternate.cls file itself is chock-full of succinct
and helpful comments.  If you consider yourself a moderately
experienced to expert user of \LaTeX, you may find reading
it useful but please remember not to change it.
}
\end{document}